\documentclass{article}

\usepackage{epsfig}

\newcommand{\screendump}[2]{
  \begingroup                
  \def\epsfsize##1##2{#1##1} 
  \epsfbox{#2}               
  \endgroup
}

\newtheorem{definition}{Definition}
\newtheorem{theorem}{Theorem}

\begin{document}

\title{Coordination Level Modeling and Analysis of Parallel Programs using Petri Nets}

\author{
        Francisco Heron de Carvalho-Junior \\
        Departamento de Computa\c{c}\~ao\\
        Universidade Federal do Cear\'a (UFC)\\
        Fortaleza, \underline{Brazil}
            \and
        Rafael Dueire Lins\\
        Centro de Inform\'atica\\
        Universidade Federal de Pernambuco\\
        Recife, \underline{Brazil}
}

\date{March 2005}

\maketitle

\begin{abstract}

In the last fifteen years, the high performance computing (HPC)
community has claimed for parallel programming environments that
reconciles generality, higher level of abstraction,
portability, and efficiency for distributed-memory parallel computing platforms.
The \textsf{Hash} component model appears as an alternative for addressing HPC
community claims for fitting these requirements. This paper
presents foundations that will enable a parallel programming
environment based on the \textsf{Hash} model to address the problems of
``debugging'', performance evaluation and verification of formal
properties of parallel program by means of a powerful, simple, and
widely adopted formalism: Petri nets.

\end{abstract}


\section{Introduction}
\label{sec:introduction}

Haskell$\#$ is a parallel extension to Haskell \cite{Carvalho2003c}, the most widely used non-strict pure functional programming language \cite{Thompson96}. It makes possible the coordination of a set of functional processes written in Haskell through a configuration language, called HCL (Haskell$\#$ Configuration Language). Thus, Haskell$_\#$ separate the programming task in two levels: the \emph{computation} level, where functional processes are written in Haskell, and the \emph{coordination} level, where functional processes are coordinated. The coordination media of Haskell$_\#$ is also called the \textsf{Hash} component model. In recent works, we have generalized the \textsf{Hash} component model in order to support other programming languages than Haskell at the computation level. Functional processes are now units, possibly written in any programming language supported by a programming environment that complies to the \textsf{Hash} component model. In this paper, we still suppose that units are the functional processes of Haskell$\#$.


The coordination level of Haskell$_\#$, i. e. the \textsf{Hash} component model, was designed in order to make possible to translate the coordination media of Haskell$\#$ programs onto Petri nets. This paper addresses the issue of
presenting a translation schema for that,
demonstrating its use to formal analysis of them, involving
verification of formal properties. The use of Petri nets allows for reusing existing
automatic tools based on this formalisms for reasoning about \textsf{Hash}
programs, such as PEP \cite{Best1997} and INA \cite{Roch99}. Further extensions could make possible performance evaluation using timed or stochastic Petri Nets variants \cite{German1997, Zimmermann2000}.

In the rest of the paper, we use the \textsf{Hash} component model to refer to the coordination level of Haskell$_\#$, i.e. it is assumed that functional processes are written in Haskell (functional modules) and they communicate through either stream-based or singleton communication channels that link their input and output ports. Such concepts are not present in the general definition of the \textsf{Hash} component model.

In additional to this introduction, this paper comprises the
following sections. Section \ref{sec:hash_model} presents
additional details about the \textsf{Hash} component model. Section
\ref{sec:translation} presents a translation schema of \textsf{Hash} programs
onto Petri nets. Section \ref{sec:formal_properties_analysis}
demonstrates how Petri net models of \textsf{Hash} programs may be used for
verification of formal concurrency properties of \textsf{Hash} programs.

\section{The \textsf{Hash} Component Model}
\label{sec:hash_model}

Distributed parallel programs may be viewed as collections of
processes that interact by exchanging messages during execution.
Current programming models provide the ability to describe
computation of processes by augmenting common languages with
notations for explicit message passing. However, they do not
provide ability to modularize concerns that appear in the design
of parallel applications, including the concern of parallelism
itself, which are scattered across the implementation of
processes. We advocate that this is the key feature for
integrating advanced software engineering techniques in the
development environment of HPC parallel applications. In
sequential programming, the focus is on modularization of
concerns, since there are a unique conceptual ``process'' and
efficiency requirements are less restrictive. This is the
essential difference that makes sequential programming actually
more suitable for current software engineering techniques for
large scale applications than current parallel programming.

The \textsf{Hash} component model may be viewed as a new paradigm for developing
message passing programs. Now, they may be viewed from two
orthogonal perspective dimensions: the dimension of
\emph{processes} and the dimension of \emph{components}.

A \textbf{process} correspond to the related notion derived from
conventional message passing programming. Thus, they are agents
that perform computational tasks, communicating through
communication channels. Conceptually, \textsf{Hash} channels, like in OCCAM
\cite{Inmos84}, are point-to-point, synchronous, typed and
unidirectional. Bounded buffers are also supported. The
disciplined use of channels is the feature that makes possible
formal analysis of parallel programs by using Petri nets, the main
topic of this paper.

A \textbf{component} is an abstract entity that address a
functional or non-functional concern of the application or its
execution environment in the parallel program. A component
describe the role of a set of processes with respect to a given
concern. The sets of components that respectively implement a set
of concerns may overlap, allowing for modular separation of
concerns that are interlaced across implementation of processes
(cross-cutting concerns). The separation of cross-cutting concerns
is an active research area in object-oriented programming of large
scale applications \cite{Kiczales1997}. A \textsf{Hash} program is defined by
a \emph{main} component that address the overall application concern. Some common examples of
cross-cutting non-functional concerns that appears in HPC
applications are: placement of processes onto processors, secure
policies for accessing computing resources on grids,
fault-tolerance schemes for long-running applications, parallel
debugging, execution timing, and so on.

\textsf{Hash}
programming is performed from the perspective of
components, instead of processes, but resulting in the specification of the topology of a network of parallel processes.

Components may be \emph{composed} or \emph{simple}. Composed
components are programmed using the \textsf{Hash} configuration language
(HCL), being built by hierarchical composition of other components, called \emph{inner components}. HCL may be viewed as
a language for gluing and orchestrating components, i.e. a connector language. It
is distinguished from other compositional languages because it
supports composition of parallel components by overlapping the concerns they address. Conventional compositional languages only allows nested
composition of sequential components. Simple components addresses functional
concerns, implemented using a host language, supposed to be
sequential. Simple components are the atoms of functionality in \textsf{Hash}
programs, constituting the leaves in their component hierarchy.

The \textsf{Hash} component model supports parallel programming with skeletons \cite{Cole1989} without any additional language support. \emph{Partial
topological skeletons} may expose topological patterns of
interaction between processes in a \textsf{Hash} program, which may be used
to produce more efficient code for specific architectures and
execution environments \cite{Carvalho2003a}. They are implemented as composed
components parameterized by their addressed concerns.

The \textsf{Hash} component model has origins in Haskell$_\#$ \cite{Carvalho2003c}, where the host language used to program simple components is Haskell. Haskell enables separation between coordination and computation code, by attaching lazy streams to communication channels at coordination level, avoiding the use of communication primitives in communication code \cite{Carvalho2002a}. This paper on focused in Haskell$_\#$.

In the next section, it is described how composed components, and
skeletons, are programmed using \textsf{Hash} configurations, while
programming of simple components, in Haskell, is described in Section
\ref{sec:simple_components}.

\begin{figure}
\centering
\includegraphics[width=1.0\textwidth]{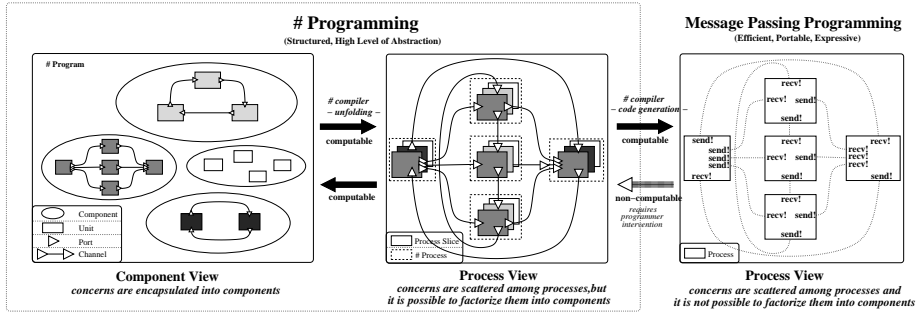}
\caption{Component Perspective versus Process Perspective}
\label{fig:component_view}
\end{figure}


\subsection{Programming Composed Components}
\label{sec:composed_components}

Composed components define coordination media of Haskell$_\#$ programs,
where all parallelism concerns are addresses without mention to
entities at computation levels, where computations are specified.
Composed components are written in HCL (\textsf{Hash} Configuration Language). Also, they define the core of the \textsf{Hash} component model, supported by Haskell$_\#$.


\begin{figure}[b]
\centering
\includegraphics[width=.8\textwidth]{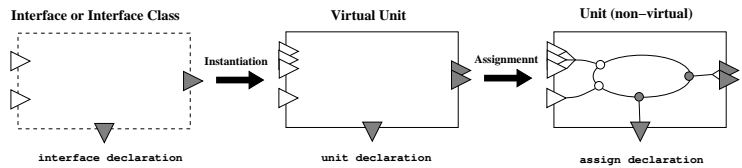}
\caption{Instantiating and Configuring a Unit} \label{fig:unit}
\end{figure}

The configuration of a composed component specifies how a
collection units, agents that perform specific tasks, interact by
means of point-to-point, typed, and unidirectional communication
channels, for addressing a given parallel programming concern. For
that, a unit is instantiated from an \emph{interface} and
associated to a \emph{component} (Figure \ref{fig:unit}). The
latter specifies the task performed by the unit, since functional modules
describe addressed concerns, while the former specifies how the
unit interacts with the coordination medium. A unit associated
with a simple component is called \emph{process}, while a unit
associated with a composed component is called a \emph{cluster}. A
interface of a unit is defined by a set of typed input and output
\emph{ports} and a \emph{protocol}. The protocol of an interface
specifies the order in which ports may be activated during
execution of the units instantiated from that interface, by means
of an embedded language whose constructors have semantic
equivalence to regular expressions controlled by semaphores. This
formalism is equivalent to place/transition Petri nets, allowing
for formal property analysis, simulation and performance
evaluation of programs using available Petri net tools
\cite{Carvalho2002b}, such as PEP \cite{Grahlmann1996} and INA
\cite{Roch99}. A port is activated between the
time in which it becomes ready to perform a communication
operation and the time it completes this operation, according to the
communication mode of the channel where the port is connected. 


\begin{figure}
\begin{center}
\includegraphics[width=0.9\textwidth]{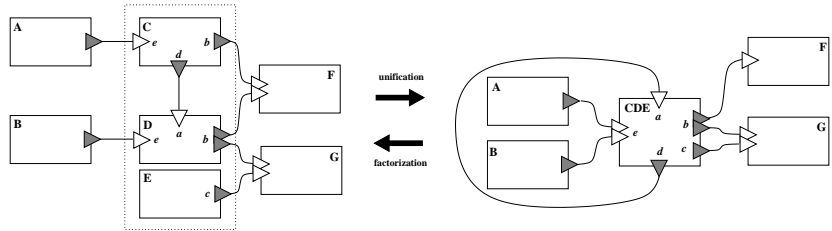}
\caption{Unification/Factorization}
\label{fig:unification_factorization_example}
\end{center}
\end{figure}

\begin{figure}[b]
\begin{center}
\includegraphics[scale=0.35,trim=0 0 0 0,clip=]{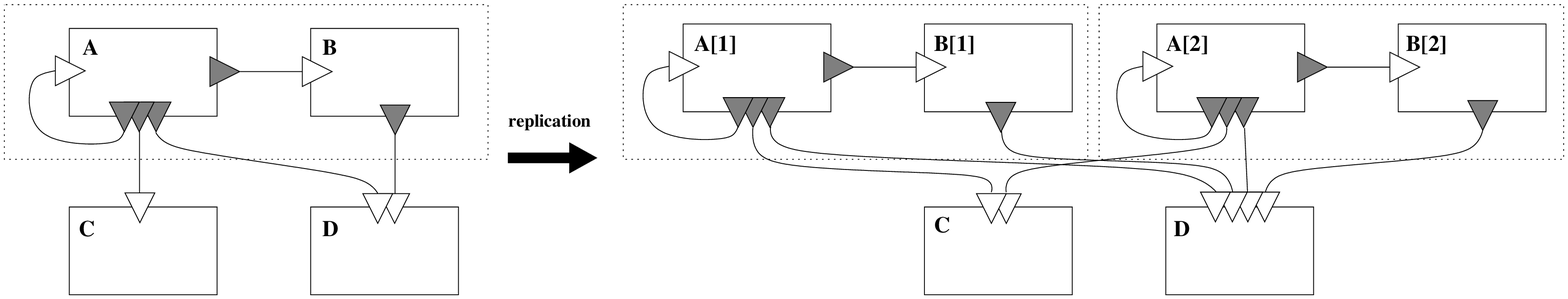}
\caption{An Illustrative Example of Replication}
\label{fig:replication_example}
\end{center}
\end{figure}

Units interact through communication channels, which connect an
output port from an unit (\emph{transmitter}) to an input port of
another one (\emph{receiver}). The types of the connected ports
must be the same. The supported communication modes, inspired on
MPI, are : \textbf{synchronous}, \textbf{buffered} and
\textbf{ready}.

In a unit specification, interface ports can be replicated to form
groups. Groups may be of two kinds: \emph{any} and \emph{all},
according to the semantics of activation. The activation of a
group of ports of kind \emph{all} implies activation of all its
port members. The activation of a group of ports of kind
\emph{any} implies that one of the port members will be put ready
for communication but only one will complete communication. The
chosen port is one of the activated ports whose communication
pairs are also activated at that instant. From the internal
perspective of the unit, groups are treated as indivisible
entities, while from the perspective of the coordination medium,
port members are referred directly in order to forming channels.
Input and output ports (groups individually) of the unit interface
must be mapped to arguments and return points of the component
assigned to it, respectively. Wire functions are useful when it is
necessary to transform values at the boundary between ports and
arguments/exit points. A particularly useful use of wire functions
is to aggregate data received from input ports belonging to a
group of ports of kind \emph{all} to a unique value, passed to the
associated argument. Similarly, wire functions allow that a value
produced in an exit point to be mapped onto a collection of values
in order to be sent by the port members of the associated group of
output ports of kind \emph{all}. Wire functions increases the
changes for reusing a component, resolving possible conflicts.

\begin{figure}
\begin{center}
\begin{minipage}{\textwidth}
\centering
\includegraphics[width=1.0\textwidth]{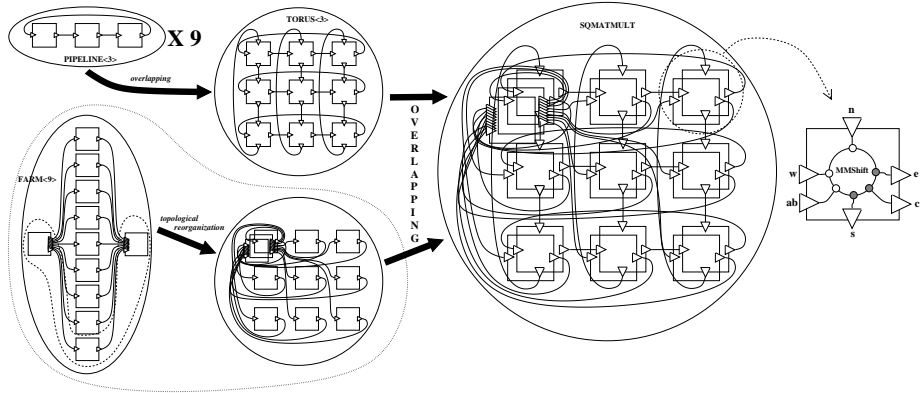}
\end{minipage}
\caption{Topology of Matrix Multiplication Using a Torus}
\label{fig:mm_topology}
\end{center}
\end{figure}

Two operations are defined on units: \emph{unification} and
\emph{factorization}. Unification allows to unify a collection of
units, forming a single unit. Factorization is the inverse of
unification, allowing units to be divided into many virtual units.
Replication, a third operation applied to units, allows that the
network induced by a collection of units to be replicated. All
operations assume that units are fully connected. Behavioral and
connectivity preserving restrictions are applied, but not
formalized here. Connectivity restrictions imply the possibility
of to replicate ports whenever it is necessary to adjust
topological connectivity after an operation. Figures
\ref{fig:unification_factorization_example} and
\ref{fig:replication_example} present illustrative examples of
these operations.

\subsubsection{Virtual Units and Skeletons}

To allow overlapping of components and the support for skeletons,
the notion of virtual unit has been introduced. A unit is virtual
whenever there is no component associated with it. In other terms,
the task performed by a virtual unit is not defined. Components
are partially parameterized in its addressed concern by means of
placing virtual units in its constitution. A component that
comprises at least one virtual unit is called an \emph{abstract
component}, or a \emph{partial topological skeleton}. These terms
are used as synonyms. When a programmer re-use an abstract
component for specification of a \textsf{Hash} program, it must to assign
components to the virtual units comprising it.

Abstract components may not instantiate applications. It is
necessary to describe the computation performed by their
constituent virtual units. The \emph{assignment} operation is used
allows to associate a component to a virtual unit, making it a
non-virtual unit. Also, there is a \emph{superseding} operation,
which allows to take a non-virtual unit for replacing a virtual
unit of the topology. The behavioural compatibility restrictions
from the non-virtual unit to the replaced virtual unit guarantees
that any sequence of communication actions that is valid in the
non-virtual unit remains valid in the virtual unit. The
superseding operation is a ``syntactic sugar'' of HCL, since it
may be implemented using unification and assignment.

\begin{figure}
\centering
\includegraphics[width=.7\textwidth]{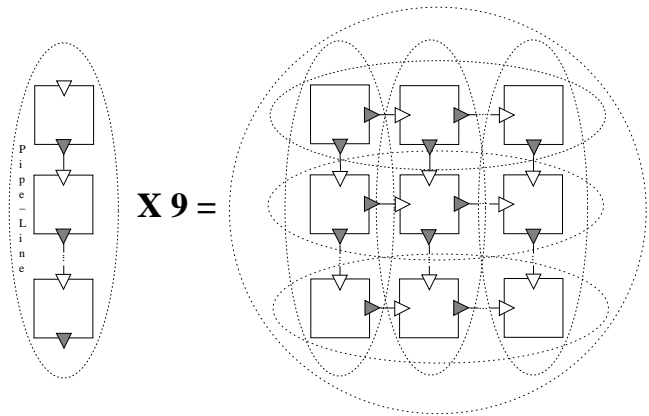}
\caption{Composing Pipe-Lines to Form a Systolic Mesh}
\label{fig:pipe_to_mesh}
\end{figure}

Figure \ref{fig:pipe_to_mesh} shows how a skeleton describing a
\emph{systolic mesh} of processes, implemented by overlapping a
collection of \emph{pipe-line} skeleton instances. The abstract
components \textsc{PipeLine} is used to describe interaction of
processes placed at the mesh lines and columns. Each unit in
\textsc{SystolicMesh} abstract component is formed by two slices:
one described by the unit that comes from the vertical
\textsc{PipeLine} component and the other described by the unit
that comes from the horizontal one.

\begin{figure}
\begin{center}
\begin{tiny}
\begin{tabular}{cc}
\begin{minipage} {\textwidth}
\begin{tabbing}

\textbf{component} \textsc{CPipeLine} $<$N$>$ \textbf{with} \\
\\
\textbf{iterator} i \textbf{range} [1,N] \\
\\
\textbf{interface} \= \emph{ICPipe} \textbf{where} \\
                   \> \textbf{ports}: i* $\rightarrow$ o* \\
                   \> \textbf{protocol}: \= \textbf{repeat} \textbf{seq}\{o!; i?\} \textbf{until} $<$o \& i$>$\\
\\
$[/$ \textbf{unit} pipe[i] \textbf{where} \textbf{ports}: \emph{ICPipe} $/]$ \\
\\
\textbf{connect} pipe[i]$\rightarrow$o \textbf{to} pipe[i+1]$\leftarrow$i, \textbf{buffered} \\
\\
\textbf{component} \textsc{Torus} $<$N$>$ \textbf{with} \\
\\
\textbf{use} Skeletons.Common.\textsc{CPipeLine} \\
\\
\textbf{iterator} i, j \textbf{range} [1,N] \\
\\
\textbf{interface} \= \emph{ITorus} \textbf{where} \\
                   \>   \textbf{ports}: \= \textit{ICPipe} @ n $\rightarrow$ s \# \textit{ICPipe} @ e $\rightarrow$ w  \\
                   \>   \textbf{protocol}: \= \textbf{repeat} \textbf{seq} \=\{\=\textbf{par} \{s!; w!\}; \textbf{par} \{n?; e?\}\} \\
                   \>                      \> \textbf{until} $<$n \& e \& s \& w$>$ \\
\\
$[/$ \textbf{unit} vpipe[i]; \textbf{assign} \textsc{CPipeLine}$<$N$>$ \textbf{to} vpipe[i] $/]$ \\
$[/$ \textbf{unit} hpipe[j]; \textbf{assign} \textsc{CPipeLine}$<$N$>$ \textbf{to} hpipe[j] $/]$ \\
\\
$[/$ \textbf{unify} \= vpipe[i].pipe[j], hpile[j].pipe[i] \\
                    \> \textbf{to} \= node[i][j] \textbf{where} \textbf{ports}: \emph{ITorus} $/]$ \\
\\
\textbf{component} \textsc{Farm}$<$N$>$ \textbf{with} \\
\\
\textbf{unit} distributor \textbf{where} \textbf{ports}: () $\rightarrow$ job \\
\textbf{unit} worker \textbf{where} \= \textbf{ports}: job $\rightarrow$ result \\
                                    \> \textbf{protocol}: \textbf{seq} \{job?; result!\}\\
\textbf{unit} collector \textbf{where} \textbf{ports}: result $\rightarrow$ () \\
\\
\textbf{connect} distributor.job \textbf{to} worker.job, \textbf{synchronous}\\
\textbf{connect} worker.result \textbf{to} collector.result, \textbf{synchronous} \\
\\
\textbf{replicate} N: worker 

\end{tabbing}
\end{minipage} &
\begin{minipage}{\textwidth}
\begin{tabbing}

\textbf{component} \textsc{SqMatMult}$<$N$>$ \textbf{where}\\
\\
\textbf{iterator} $i$, $j$ \textbf{range} [1,N] \\
\\
\textbf{use} Skeletons.Common.\{\textsc{Torus}, \textsc{Farm}\} \\
\textbf{use} \textsc{MMShift} \\
\\
\textbf{int}\=\textbf{erface} \emph{ISqMatMult} \textbf{where} \\
                   \> \textbf{ports}: j $\rightarrow$ r \# ITorus \\
                   \> \textbf{protocol}: \textbf{seq} \{j?; \= \textbf{repeat} \textbf{seq} \{s!;e!;n?;w?\} \\
                   \>                                       \> \textbf{counter} $N$ \} \\
\\
\textbf{unit} mm\_torus; \textbf{assign} \textsc{Torus}$<$N$>$ \textbf{to} mm\_torus \\
\textbf{unit} mm\_farm; \textbf{assign} \textsc{Farm}$<$N$>$ \textbf{to} mm\_farm \\
\\
$[/$ \textbf{unify} \= farm.worker[$i + j \times N$], torus.node[$i$][$j$] \\
                    \> \textbf{to} sqmm[$i$][$j$] $/]$ \textbf{where} \textbf{ports}: \emph{ISqMatMult} \\
\\
\textbf{unify} \= farm.distributor, farm.collector, sqmm[0][0] \\
               \> \textbf{to} \= sqmm\_root \textbf{where} \\
               \>             \> \textbf{ports}: \= () $\rightarrow$ ab \# c $\rightarrow$ () \# \\
               \>             \>                 \> \emph{ISqMatMult} @ mm \\
               \>             \> \textbf{protocol}: \textbf{seq} \{ab!; c?; \textbf{do} mm \}\\
\\
$[/$ \textbf{assign} \textsc{MMShift} to sqmm[$i$][$j$] $/]$ \\
\\
\textbf{module} \textsc{MMShift}(\emph{main}) \textbf{where} \\
\\
\emph{main} :: Num t $\Rightarrow$ t $\rightarrow$ t $\rightarrow$ [t] $\rightarrow$ [t] $\rightarrow$ ([t],[t],t) \\
\emph{main} \= a b as\_i bs\_i = (as\_o,bs\_o,c) \\
            \> \textbf{where} \= \\
            \>                \> c = \emph{matmult} as\_o bs\_o \\
            \>                \> (as\_o, bs\_o) = (a:as\_i, b:bs\_i) \\
\\
\emph{matmult} :: Num t $\Rightarrow$ t $\rightarrow$ [t] $\rightarrow$ [t] $\rightarrow$ t \\
\emph{matmult} [] [] = 0 \\
\emph{matmult} (a:as) (b:bs) = a*b + \emph{matmult} as bs

\end{tabbing}
\end{minipage}
\end{tabular}
\end{tiny}
\caption{Configuration Code of Matrix Multiplication on a Torus}
\label{fig:mm_configuration_code}
\end{center}
\end{figure}

\begin{figure}
\begin{center}
\begin{footnotesize}
\begin{minipage} {\textwidth}
\begin{tabbing}

\textbf{module} Tracking(\emph{main}) \textbf{where} \\
\\
\textbf{import} Track \\
\textbf{import} Tallies \\
\textbf{import} Mcp\_types \\
\\
\emph{main} :: User\_spec\_info $\rightarrow$ [(Particle,Seed)] $\rightarrow$ ([[Event]],[Int]) \\
\emph{main} \= user\_info particle\_list = \= \textbf{let} \= events's = map f particle\_list  \textbf{in} (events's, tally\_bal event\_lists) \\
            \>  \textbf{where} \= \\
            \>                 \> f (particle@(\_,\_,\_, e, \_), sd) = (Create\_source e):(track user\_info particle [] sd) 

\end{tabbing}
\end{minipage}
\end{footnotesize}

\caption{A Functional Module from MCP-Haskell$_\#$}
\label{fig:tracking_functional_module}
\end{center}

\end{figure}

\subsection{Programming Simple Components}
\label{sec:simple_components}

Simple components, also called \emph{functional modules}, are
atoms of functionalities in \textsf{Hash} programming. The collection of
simple components in a \textsf{Hash} program describes its computation media.
Simple components might be programmed virtually in any general
purpose language, called \emph{host language}. For that, it is
needed to define which host language constructions correspond to
the arguments and return points of the underlying functional
module. It is preferred that no extensions to the host language be
necessary for this purpose, keeping transparency between
coordination and computation media. Simple components may be
overlapped when configuring composed components. The goal is to
implement a really multi-lingual approach for parallel
programming. For that, it has been proposed to use CCA (Common
Component Architecture) \cite{Armstrong1999}, a recent standard
proposed for integrating components written in different languages
in a parallel environment. Another possibility is to use
heterogeneous implementations of MPI \cite{Squyres2000,
Squyres2003a}, recently proposed, facilitating this task since, in
cluster environments, \textsf{Hash} programs are compiled to MPI.

Since the translation schema onto Petri nets is defined on top of
coordination media, abstracting from computation media concerns,
specific details about programming simple components is not
provided in this paper. For illustration, Figure
\ref{fig:tracking_functional_module} presents an example of
functional module from MCP-Haskell$_\#$ program
\cite{Carvalho2001a}, written in Haskell.

\begin{figure}
\begin{center}
\begin{minipage}{\textwidth}
\begin{tiny}
\begin{tabbing}

$<$\textsc{Component}$>$ $\rightarrow$ \textbf{component} (\=\{$u_1$:$<$\textsc{Unit}$>$, $\dots$, $u_{i}$:$<$\textsc{unit}$>$\}, \{$c_1$:$<$\textsc{Channel}$>$, $\dots$,$c_{j}$:$<$\textsc{Channel}$>$\}) \\
\\
$<$\textsc{Unit}$>$ $\rightarrow$ \textbf{unit} ($id$, \{$p_1$:$<$\textsc{Port}$>$, $\dots$, $p_{k}$:$<$\textsc{Port}$>$\}, $<$\textsc{Type$_u$}$>$)\\
\\
$<$\textsc{Type$_u$}$>$ $\rightarrow$ \textbf{repetitive} $\mid$ \textbf{non-repetitive} \\
\\
$<$\textsc{Behavior}$>$ $\rightarrow$ \textbf{protocol} ($\{id^{sem}_1,\dots,id^{sem}_n\}$, $<$\textsc{Action}$>$) \\
\\
$<$\textsc{Action}$>$ \= $\rightarrow$      \= \textbf{skip} \\
                \> $\mid$ \> \textbf{seq} \{$a_1$:$<$\textsc{Action}$>$; $\dots$; $a_k$:$<$\textsc{Action}$>$\} \\
                \> $\mid$ \> \textbf{par} \{$a_1$:$<$\textsc{Action}$>$; $\dots$; $a_k$:$<$\textsc{Action}$>$\} \\
                \> $\mid$ \> \textbf{alt} \{$g_1$:$<$\textsc{Action}$>$; $\dots$; $g_k$:$<$\textsc{Action}$>$\} \\
                \> $\mid$ \> \textbf{repeat\_until} ($<$\textsc{Action}$>$, C) \\
                \> $\mid$ \> \textbf{repeat\_counter} ($<$\textsc{Action}$>$, N) \\
                \> $\mid$ \> \textbf{repeat\_forever} $<$\textsc{Action}$>$ \\
                \> $\mid$ \> \textbf{signal} $id$\\
                \> $\mid$ \> \textbf{wait} $id$\\
                \> $\mid$ \> \textbf{activate} $id$ \\
\\
$<$\textsc{Port}$>$ $\rightarrow$ \textbf{port}  ($id$,$<$\textsc{Direction}$>$,$<$\textsc{Multiplicity}$>$,$<$\textsc{Type$_p$}$>$,$nesting\_factor$)\\
\\
$<$\textsc{Multiplicity}$>$ \= $\rightarrow$ \= \textbf{single} $\mid$ \textbf{group} ($<$\textsc{Type$_g$}$>$, \{$p_1$:$<$\textsc{Port}$>$, $\dots$, $p_n$:$<$\textsc{Port}$>$\}) \\
\\
$<$\textsc{Direction}$>$ $\rightarrow$ \textbf{input} $\mid$ \textbf{output}\\
\\
$<$\textsc{Type$_p$}$>$ $\rightarrow$ \textbf{stream} $\mid$ \textbf{non-stream}\\
\\
$<$\textsc{Type$_g$}$>$ $\rightarrow$ \textbf{any} $\mid$ \textbf{all} \\
\\
$<$\textsc{Channel}$>$ $\rightarrow$ \textbf{connect} ($id^{port}$, $id^{port}$, $<$\textsc{ChanMode}$>$) \\
\\
$<$\textsc{ChanMode}$>$ $\rightarrow$ \textbf{synchronous} $\mid$ \textbf{buffered} $\mid$ \textbf{ready} 

\end{tabbing}
\end{tiny}
\end{minipage}
\end{center}

\label{abstract_hash} \caption{Abstract \textsf{Hash} Configuration Language
Syntax}

\end{figure}

\subsection{An Abstract Representation for \textsf{Hash} Components}

In Figure \ref{abstract_hash}, it is defined a simplified syntax
for an abstract representation of \textsf{Hash} configurations, named
\emph{Abstract \textsf{Hash}}. The abstract \textsf{Hash} configuration language (AHCL)
only captures information strictly relevant to the translation
schema onto Petri nets further presented. For example, interface
declarations and operations over units, such as unifications,
factorizations, replications, and assign operation are not
represented in AHCL. It is supposed that these operations are all
resolved before translation process. AHCL is not a simplification.
Indeed, in the compilation process of \textsf{Hash} configurations, AHCL
correspond to the intermediate code generated by the
\emph{front-end} compiler module, which serves as input to all
\emph{back-end} modules. A \emph{back-end} was developed for
generating PNML (Petri Net Markup Language) code from a \textsf{Hash}
configuration. The following paragraph describes the structure of
AHCL.

A component is composed by a set of units ($u_1, \dots, u_i$) and
a set of channels ($c_1, \dots, c_j$). Units are described by an
identifier, a collection of ports ($<\textsc{Port}>$). A unit can
be \textbf{repetitive} or \textbf{non-repetitive}
($<\textsc{Type$_u$}>$). The interface of a unit is defined by a
collection of ports and a protocol, described by means of an
embedded language that specifies valid orders for activation of
ports. This language have constructors equivalent to combinators
of regular expressions controlled by balanced semaphores. A port
is described by an identifier ($id$), a direction
($<\textsc{Direction}>$), a type (stream or non-stream) and a
nesting factor. Additionally, port \emph{multiplicity} specifies
if a port is a single port or a group of ports. Notice that a
group of port can be of two types: \textbf{any} or \textbf{all}.
All port identifiers are assumed to be distinct in an abstract \textsf{Hash}
programs. A channel connect two ports and is associated to a mode
(\emph{synchronous}, \emph{buffered} and \emph{ready}). Notice
that, abstract \textsf{Hash} syntax does not force that these two ports be
from opposite directions. This restriction is implicitly assumed.

\begin{figure}
\begin{center}
\begin{minipage}{\textwidth}
    \centering
    \screendump{0.8}{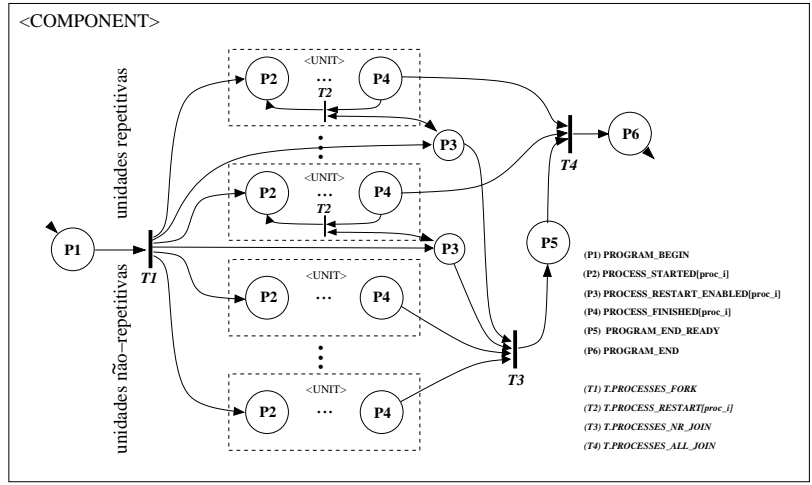}
\end{minipage}

\end{center}

\caption{Translating a Component}

\label{fig:translation_component}
\end{figure}

\section{Translating \textsf{Hash} programs into Petri Nets}
\label{sec:translation}

In this section, a schema for translating \textsf{Hash} programs into Petri
nets is introduced. In order to make translation schema easier to
understand, it is informally described using diagrams. The
possibility of making intuitive visual descriptions is an
interesting feature of Petri nets. 

The translation schema is specified inductively in the hierarchy
of components at the coordination medium of the \textsf{Hash} program. Thus,
simple components are ignored. The overall steps in the
translation procedure of a \textsf{Hash} program into an interlaced Petri net
are:

\begin{enumerate}

\item \textbf{Translating units}: For each unit comprising the
component (\textbf{unit} declarations), its interface is used for
yielding an interlaced Petri net describing activation order of
their interface ports. In \textsf{Hash} configuration language, it is defined
by an embedded language, in interface declarations, whose
combinators have correspondence to operators of regular expression
controlled by semaphores. This formalism was proved to have
expressiveness equivalence to Petri nets according to formal
language theory. If a unit is a cluster (a composed component is
assigned to the unit), it is necessary to generate the Petri net
that corresponds to the assigned component. Using information
about mapping of argument/return points to input/output ports of
the unit, it is possible to synchronize behavior of the unit with
behavior of the component, in such way that they are compatible;

\item \textbf{Synchronize units}. Now that a Petri net exists for
describing communication behavior (traces) of each unit,
communication channels (\textbf{connect} declarations) may be used
for coordinating synchronized behavior of these Petri nets
(units);

\item \textbf{Synchronize streams}. For each port carrying a
stream, a Petri net describing a protocol for stream
synchronization is overlapped with the Petri net produced in the
last step. The semantics of stream communication is described
separately because it increases complexity of the generated
interlaced Petri net, making computationally hard its analysis.
Thus, in \textsf{Hash} programming environment, the programmer may decide not
to include stream synchronization protocol. Obviously, information
may be lost, but it may not be necessary for some useful analysis.

\end{enumerate}

The next sections provide details about the above translation
steps. Also, it is discussed how higher-level information
encompassed in skeletons may be used to simplify the generated
network.

\begin{figure}
\begin{center}
\begin{minipage}{\textwidth}
\centering
\includegraphics[scale=0.45,trim=0 0 0 0,clip=]{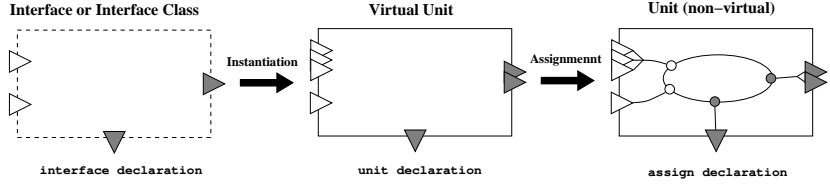}
\end{minipage}
\caption{Unit = Interface (by instantiation) + Component (by
assignment)} \label{fig:assignment_figure}
\end{center}
\end{figure}

\subsection{Modelling Components $(\Upsilon^C)$}
\label{sec:translation_component}

In Figure \ref{fig:translation_component}, the Petri net resulted
from application of translation function $\Upsilon^C$ to the
configuration of a component is illustrated. The translation
function $\Upsilon^U$ is applied to each unit comprising the
component, generating a Petri net that describes its communication
behavior. The resulting Petri nets are connected in order to model
parallel execution of units. The places
\textsc{process\_started}[$j$] and
\textsc{process\_finished}[$j$], $1 \leq j \leq n$, where $n$ is
the number of units, correspond to start places and stop places of
the Petri nets modelling units, respectively. When a token is
placed on \textsc{process\_started}[$j$], the unit $j$ is ready to
initiate execution, and when a token is placed on
\textsc{process\_finished}[$j$], the unit $j$ had finished. The
transitions \textsc{process\_restart}[$k$], $1 \leq k \leq r$,
where $r$ is the number of repetitive units, allows that
repetitive units return back to its initial state after
finalization. They are introduced in the Petri net generated for
each unit. The place \textsc{program\_end\_ready} receives a mark
when all non-repetitive units terminates. In this case, the tokens
in places \textsc{process\_restart\_enabled}[$k$], $1 \leq k \leq
r$, are removed, preventing repetitive processes execute again. At
this state, the program terminates after all repetitive processes
also terminate, which causes transition
\textsc{processes\_all\_join} to be fired and a token to be
deposited in \textsc{program\_end}.

\begin{figure}
\begin{center}
\begin{minipage}{\textwidth}
    \centering
    \screendump{0.4}{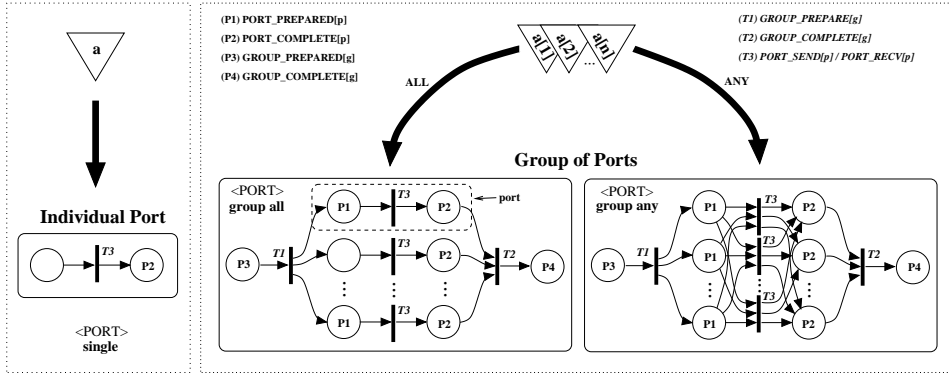}
\end{minipage}

\end{center}

\caption{Translating Unit's Ports}

\label{fig:unit_ports_diagram}
\end{figure}

\subsection{Modelling Units $(\Upsilon^U)$}
\label{sec:modelling_units}

This section intends to describe how individual units are
translated into Petri nets. Firstly, Figure
\ref{fig:unit_ports_diagram} presents interlaced Petri nets that
model the activation of the ports of the interface of a unit. A
mark in place \textsc{port\_prepared}[$p$] indicates that the port
$p$ is prepared for communication. The firing of transition
\textsc{port\_send}[$p$]/\textsc{port\_recv}[$p$] models
communication, causing the deposit of a mark in place
\textsc{port\_complete}[$p$] for indicating that communication has
been completed on port $p$. It is possible that two ports be
active at the same time. In groups of ports, the places
\textsc{group\_prepare}[$g$] and \textsc{group\_complete}[$g$] are
connected to places \textsc{port\_prepared}[$p$] and
\textsc{port\_complete}[$p$] of each port $p$ belonging to the
group $g$ that model local preparation and completion of
communication in ports belonging to the group, according to their
semantics: \emph{any} or \emph{all}. In groups of ports of kind
\emph{all}, all individual ports are activated in consequence of
activation of the group. In groups of ports of kind \emph{any},
only one port is chosen among the ports that are ready for
communication completion. For obeying this semantic restriction,
the firing of a transition
\textsc{port\_send}[$p$]/\textsc{port\_recv}[$p$] of a port $p$ in
group $g$ of kind \emph{any} causes removal of the marks in places
\textsc{port\_prepared}[$p'$], for all places $p'$ belonging to
$g$, in such a way that all ports $p$, such that $\overline{p}
\neq p$, cannot complete communication (firing of transition
\textsc{port\_send}[$\overline{p}$]/\textsc{port\_recv}[$\overline{p}$]).

The order in which marks are placed in places
\textsc{port\_prepared}[$p$], for any $p$, is controlled by an
interlaced Petri net that models the protocol of the unit. The
following sections discusses how primitive actions and action
combinators of behavior expressions are translated into Petri
nets. The primitive actions are: \textbf{skip} (null action),
\textbf{wait} (increment semaphore primitive), \textbf{signal}
(decrement semaphore primitive), \textbf{?} (activation of input
port), and \textbf{!} (activation of output ports). The
combinators of actions are: \textbf{seq} (sequential actions),
\textbf{par} (concurrent actions), \textbf{alt} (non-deterministic
choice among $n$ actions), and \textbf{if} (conditional choice
between two actions).


\begin{figure}
\begin{center}
\begin{minipage}{\textwidth}
    \centering
    \screendump{1.0}{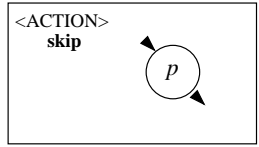}
\end{minipage}


\end{center}

\caption{Translating ``Null Action''}

\label{fig:skip_translation_diagram}
\end{figure}

\subsubsection{Null Action (skip)} \label{sec:translation_skip}

The \textbf{skip} combinator have no communication effect. Because
that, it is known as the ``null action''. Only one place is
needed, where it is both \emph{start} and \emph{stop} place
(Figure \ref{fig:skip_translation_diagram}) of the interlaced
Petri net generated.

\subsubsection{Sequencing (seq)} \label{sec:translation_seq}

\begin{figure}[b]
\begin{center}
\begin{minipage}{\textwidth}
    \centering
    \screendump{0.8}{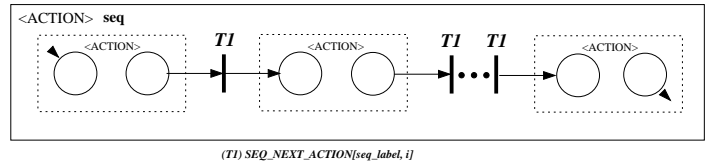}
\end{minipage}


\end{center}

\caption{Sequence of Actions}

\label{fig:sequence_translation_diagram}
\end{figure}

The \textbf{seq} combinator describes a total ordering for
execution of a set of actions (sequential execution), represented
by $a_1, a_2, \dots, a_n$. It may be modelled by sequential
composition of the Petri nets induced for each action (Figure
\ref{fig:sequence_translation_diagram}).

\begin{figure}
\begin{center}
\begin{minipage}{\textwidth}
    \centering
    \screendump{0.8}{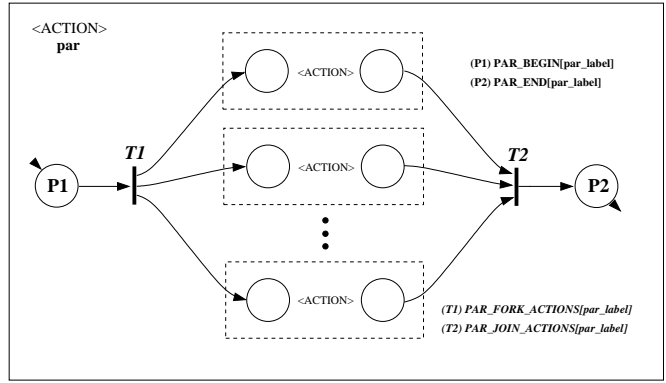}
\end{minipage}

\end{center}

\caption{Interleaving (Concurrency) Among Actions}

\label{fig:concurrency_translation_diagram}
\end{figure}

\subsubsection{Concurrency (par)} \label{sec:translation_alt}

The \textbf{par} combinator describes a concurrent (interleaving)
execution of a set of actions, represented by $a_1, a_2, \dots,
a_n$. It may be modelled by parallel composition of the Petri nets
induced by each action (Figure
\ref{fig:concurrency_translation_diagram}).

\subsubsection{Non-Deterministic Choice (alt)} \label{sec:translation_alt}

\begin{figure}[b]
\begin{center}
\begin{minipage}{\textwidth}
    \centering
    \screendump{0.8}{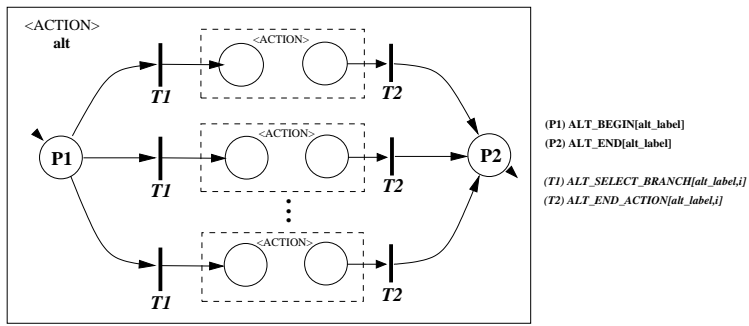}
\end{minipage}

\end{center}

\caption{Choice Among Actions}

\label{fig:alt_translation_diagram}
\end{figure}

The \textbf{alt} combinator describes a conceptually
non-deterministic choice among a set of actions, represented by
$a_1, a_2, \dots, a_n$. It may be modelled by composing the Petri
nets induced for each action using a conflict in place
\textsc{alt\_begin}[$alt\_label$], its start place (Figure
\ref{fig:alt_translation_diagram}). The firing of transitions
\textsc{alt\_select\_branch}[$alt\_label$, i], $1 \leq i \leq n$,
models the choice.

\subsubsection{Streams and Conditions for Checking Stream Termination}
\label{sec:modelling_stream_communication}

The two combinators modelled in the next sections, \textbf{repeat}
and \textbf{if}, requires testing a condition in order to choose
the next action to be performed. The condition is defined by a
logical predicate in its disjunctive normal form (DNF). Logical
variables are references to ports that carry \emph{streams}. This
section attempts to formalize the notion of streams in the \textsf{Hash} component
model and valuation of logical variables of termination-test
conditions of streams.

A stream is defined as a sequence of semantically related data
items, terminated with a special mark, that is transmitted through
a channel. By making an analogy with conventional message passing
programming using MPI, a trivial example of stream is a sequence
of data items transmitted in calls to a specific occurrence of
\texttt{MPI\_Send} primitive in the context of an iteration. At
each iteration, an item of the stream is transmitted. The
termination of the iteration is modelled, in the \textsf{Hash} program, by
the end mark carried by the stream. Communication on channels that
carry streams may be implemented using persistent communication
objects in the underlying messaging passing library, which may
reduce communication overhead.

\textsf{Hash} streams may be nested. Streams of streams, at any nesting depth
may be defined. A stream $S_{x}^{(1,d)}$, where $d$ is the nesting
factor of the stream and $x$ is a positive integer that indicates
the order of a nested stream in a stream, may be defined as
following:

\begin{equation}
\begin{footnotesize}
\begin{array} {c}

S_{x}^{(i,d)} = \left\{\begin{array}{ll} \langle S_1^{(i+1,d)},S_2^{({i+1},d)},\dots,S_n^{({i+1},d)}, \texttt{EOS}[i] \rangle, n \geq 0 & \mbox{, if $i < d$} \\
                                    \texttt{VALUE} & \mbox{, if $i = d$}
                 \end{array}
          \right.\\

\label{eq:stream_definition}
\end{array}
\end{footnotesize}
\end{equation}

where $\texttt{VALUE}$ is a data item and $\texttt{EOS}[i]$ is a
termination value at nesting level $i$. Notice that termination
values should carry an integer indicating the nesting level of the
stream being terminated. The feature of nested streams appeared in
consequence of design of Haskell$_\#$. Streams at coordination
level must be associated to lazy lists in computation media
\footnote{In Haskell$_\#$, simple components are functional
modules written in Haskell}. The experience with Haskell$_\#$
programming have shown that laziness of Haskell nested lists may
be useful in some applications. This feature is analogous to
communication operations that occurs in context of nested
iterations in MPI parallel programming.

In \textsf{Hash} configuration language, streams are declared by placing
``*'' symbols after the identifier of a port in the declaration of
\textit{interfaces}. The number of ``*'''s indicates the nesting
factor of the stream carried by the port. Only ports carrying
streams of the same nesting levels may be connected through a
channel. It is defined that a port that transmit a single value
(non-streamed) have nesting level zero.

Now that the notion of streams is defined, it is possible to
define the syntax and semantics of predicates for testing
synchronized termination of streams, a necessary feature for
combinators \textbf{if} and \textbf{repeat}. This kind of
predicate will be referred as \emph{stream predicate}.

Syntactically, a stream predicate is a logical predicate in its
disjunctive normal form. The logical operators supported are
``\&'', the logical \emph{and}, and ``$|$'', the logical
\emph{or}. Disjunctions may be enclosed by ``$\langle$'' and
``$\rangle$'' delimiters. Logical variables are references to
interface ports of a unit. The formal syntax of stream predicates
is shown below:

\begin{center}
\begin{tabbing}

\emph{stream\_predicate} \hspace*{0.3cm} \= $\rightarrow$ \emph{sync\_conjunction}$_1$ `$\mid$' $\dots$ `$\mid$' \emph{sync\_conjunction}$_n$\ \ \  ($n \geq 1$)\\
\emph{sync\_conjunction} \> $\rightarrow$ `$\langle$' \emph{simple\_conjunction} `$\rangle$' $\mid$ \emph{simple\_conjunction} \\
\emph{simple\_conjunction} \> $\rightarrow$ port\_id $\mid$ ( port\_id$_1$ `\&' $\dots$ `\&' port\_id$_n$ )\ \ \   ($n \geq 1$) 

\end{tabbing}
\end{center}

Let $d$ be the depth of nesting of an occurrence $R$ of a
\textbf{repeat} or \textbf{if} combinator in relation to an
outermost occurrence, if it exists ($d=0$, if it does not exists).
Only port carrying streams with nesting factor equal or less than
$d$ can appear in the termination condition of $R$. It is now
possible to define semantics of stream predicates, by defining how
values of logical variables may be inferred in execution of units.
For instance, let $p$ be a port carrying stream $S^{(1,n)}$. Its
value is \emph{false} whenever a data value (\texttt{VALUE}) or an
ending value at nesting level $i$ (\texttt{EOS[$i$]}), such that
$i < d$, was transmitted (sent or received) in its last
activation. Otherwise, it is \emph{true}. The value of the stream
predicate may evaluate to \textbf{true}, \textbf{false} or
\textbf{fail}. A value \textbf{true} is obtained by evaluating the
stream predicate ignoring semantics of angle brackets delimiters.
A \textbf{fail} is obtained if negation of some conjunction
enclosed in angle brackets evaluates to true, assuming the
following identity that defines angle brackets semantics:

\begin{equation}
\begin{footnotesize}
\begin{minipage} {\textwidth}
\centering

$\neg \langle a_{(k,r)} \wedge a_{(k,r+1)} \wedge \dots \wedge a_{(k,s)} \rangle = \neg a_{(k,r)} \wedge \neg a_{(k,r+1)} \wedge \dots \wedge \neg a_{(k,s)} $

\end{minipage}
\end{footnotesize}
\label{identity_sync_stream}
\end{equation}

If stream predicate evaluates neither to \textbf{true} or
\textbf{error}, the value of the stream predicate is
\textbf{false}. Angle brackets delimiters are used for ensuring
synchronization of the nature of values transmitted by streams,
whenever necessary.


In order to make possible to model test of stream predicates with
Petri nets, it is firstly necessary to model stream communication
by using this formalism.

Particularly, it is necessary to introduce, in the Petri net of
the \textsf{Hash} program, places that can remember the kind of value
transmitted in the last activation of ports that carry streams.


\begin{figure}
\begin{center}
\begin{minipage}{\textwidth}
    \centering
    \screendump{0.8}{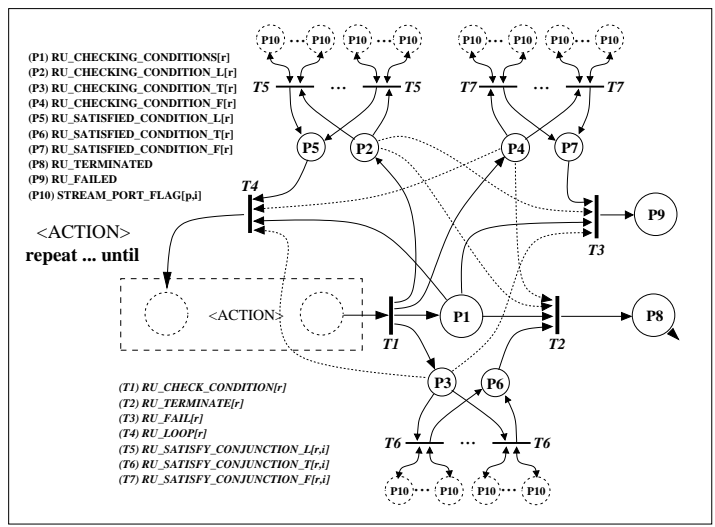}
\end{minipage}

\end{center}
\caption{Stream Controlled Repetition}

\label{fig:repeat_until_translation}
\end{figure}

In a \textsf{Hash} program, for each port $s$ that carries a stream with
nesting factor $n$, there are two sets of $n+1$ places, here
referred as $\textsc{Stream\_Flags(s)}$ and
$\overline{\textsc{Stream\_Flags}}(s)$:

\begin{footnotesize}
\[ \textsc{Stream\_Flags(s)} = \{\textsc{stream\_port\_flag}[s,k] \mid 0 \leq k \leq n \} \]
\[ \overline{\textsc{Stream\_Flags}(s)} = \{\textsc{stream\_port\_flag\_dual}[s,k] \mid 0 \leq k \leq n \}\]
\end{footnotesize}

For some stream port $s$, the places in set
$\textsc{Stream\_Flags(s)}$ form a split binary semaphore:

\begin{footnotesize}

\[ \sum_{\textsc{p} \in \textsc{Stream\_Flags(s)}} M(\textsc{p}) = 1 \]

\end{footnotesize}

Also, they are mutually exclusive with its corresponding places in
$\overline{\textsc{Stream\_Flags}}(s)$:

\begin{footnotesize}

\[ \forall k: 0 \leq k \leq n: M(\textsc{stream\_port\_flag}[s,k]) + M(\textsc{stream\_port\_flag\_dual}[s,k]) = 1 \] 

\end{footnotesize}

For a port $s$ carrying a stream with nesting factor $n$, the
places $\overline{\textsc{Stream\_Flags}}(s)$ and
$\textsc{Stream\_Flags(s)}$ are used to remember which kind of
value was transmitted in the last activation of $s$. There are
$n+1$ possibilities:

\begin{itemize}

\item{An ending marker at nesting level $i$ (\textsc{EOS[$i$]}),
for $0 \leq i \leq n-1$};

\item{A data value}.

\end{itemize}

The places \textsc{stream\_port\_flag}[$s$,$k$], such that $0 \leq
k \leq n-1$, are respectively associated to ending marks
\textsc{EOS[$k$]} of a port $s$ carrying a stream with nesting
factor $n$. The place \textsc{stream\_port\_flag}[$s$,$n$] is
associated to a data value. Assuming the restrictions above, if
there is a mark on the \textsc{stream\_port\_flag}[$s$,$k$] place,
then a value of its corresponding kind was transmitted in the last
activation of the port.

All the above restrictions are guaranteed by the Petri net
protocol for synchronization of streams introduced further, in
Section \ref{sec:stream_syncronization_protocol}. The next two
sections present respectively how to model \textbf{repeat} and
\textbf{if} combinators, assuming the existence of sets of places
${\textsc{Stream\_Flags}}(s)$ and
$\overline{\textsc{Stream\_Flags}}(s)$.

\begin{figure}
\begin{center}
\begin{minipage}{\textwidth}
    \centering
    \screendump{0.8}{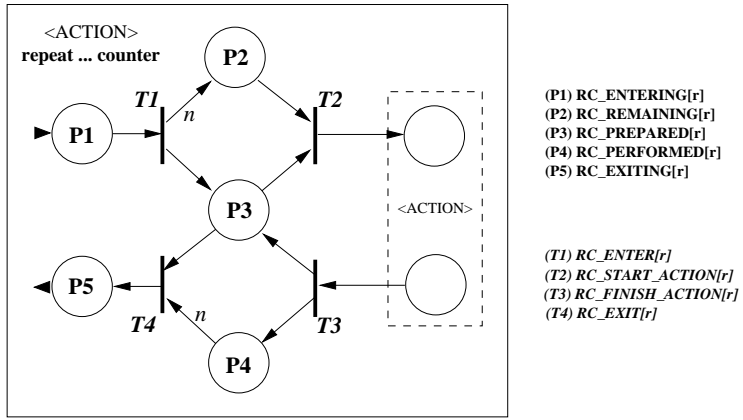}
\end{minipage}

\end{center}

\caption{Repeated Action by a Fixed Number of Times}

\label{fig:repeat_fixed_diagram}
\end{figure}

\begin{figure}
\begin{center}
\begin{minipage}{\textwidth}
    \centering
    \screendump{0.8}{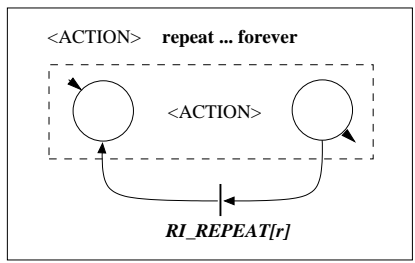}
\end{minipage}

\end{center}

\caption{Infinite Repetition (\textbf{repeat})}

\label{fig:repeat_infinite_diagram}
\end{figure}

\subsubsection{Repetition Controlled by Stream Predicates}
\label{sec:translation_unbounded_repeat}

The \textbf{repeat} combinator is used to model repeated execution
of an action. The termination condition may be provided in a
\textbf{until} or a \textbf{counter} clause. The later will be
introduced in the next section. The former uses a stream predicate
for testing termination of a repetition after each iteration. In
the Petri net modelling of \textbf{repeat} combinator with
\textbf{until} clause, it is assumed existence of the sub-network
modelling stream communication introduced in Section
\ref{sec:modelling_stream_communication}.

In Figure \ref{fig:repeat_until_translation}, it is illustrated
the Petri net resulted from translation of \textbf{repeat}
combinator with an \textbf{until} clause to check termination. The
conflict in place \textsc{ru\_checking\_conditions} models the
decision on to terminate or not the iteration. Values of logical
variables corresponding to ports in the stream predicate are
tested using their respective
$\overline{\textsc{Stream\_Flags}}(s)$ and
$\textsc{Stream\_Flags(s)}$ places. The arrangement of places and
transitions in Figure \ref{fig:repeat_until_translation} allows
for testing the value of stream predicates at each iteration. The
mutual exclusive firing of transitions \textsc{ru\_terminate},
\textsc{ru\_fail} and \textsc{ru\_loop} correspond, respectively,
to values \textbf{true} (execute one more iteration),
\textbf{false} (termination) and \textbf{error} (abort the
program) for the stream predicate.

\subsubsection{Bounded Repetition}
\label{sec:translation_bounded_repeat}

The combinator \textbf{repeat} with termination condition defined
by a \textbf{counter} clause models repeated execution of an
action by a fixed number of times. Its translation into Petri nets
is illustrated in Figure \ref{fig:repeat_fixed_diagram}. For a
certain fixed bounded repetition $r$, the weight of arcs
$(\textsc{rc\_enter}[$r$],\textsc{rc\_remaining}[$r$])$ and
$(\textsc{rc\_performed}[$r$],\textsc{rc\_exit}[$r$])$ ($n$)
define the number of repetitions of $<ACTION>$.

\subsubsection{Infinite Repetition}
\label{sec:translation_infinite_repeat}

Whenever no termination condition is defined for a given
occurrence of the \textbf{repeat} combinator, the given action is
repeated infinitely. The Petri net that models this kind of
\textbf{repeat} combinator is illustrated in Figure
\ref{fig:repeat_infinite_diagram}.

\begin{figure}
\begin{center}
\begin{minipage}{\textwidth}
    \centering
    \screendump{0.8}{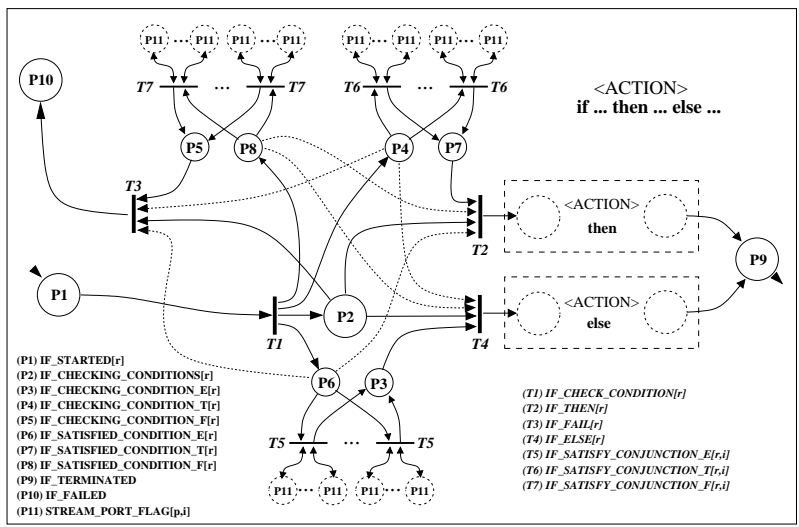}
\end{minipage}

\end{center}
\caption{Conditional}

\label{fig:if_translation}
\end{figure}

\subsubsection{Conditional Choice (if)} \label{sec:translation_if}

The \textbf{if} combinator describes a conditional choice between
to two actions. Its translation into Petri nets is illustrated in
Figure \ref{fig:if_translation}. The reader may notice the analogy
between the construction of the Petri net in this diagram with the
Petri net generated for the \textbf{repeat ... until} combinator,
mainly concerning the test of the \emph{stream predicate}.

\subsubsection{Semaphore Primitives}
\label{sec:translation_semaphores}

The \textbf{wait} and \textbf{signal} (balanced counter) semaphore
primitives and the \textbf{par} combinator make behavior
expressions comparable to labelled Petri nets in descriptive power
\cite{Ito1982}. In absence of semaphore primitives, only regular
patterns of unit behavior could be described. The semantic of
semaphore primitives is now defined using the notation introduced
in \cite{Andrews1991} for concurrent synchronization:

\begin{minipage} {\textwidth}
\centering
\begin{tabbing}
\\
\=$wait(s)$: $<${\bf await} $s > 0 \rightarrow  s := s - 1 $$>$ \\
\>$signal(s)$: $<$ $s := s$ + 1 $>$
\\
\end{tabbing}
\end{minipage}

The angle brackets model mutual exclusion (atomic actions in
concurrent processes), while the \textbf{await} statement models
condition synchronization. The \textbf{wait} primitive causes a
process to delay until the value of the semaphore $s$ is greater
than zero in order to decrement its value. Thus, the value of a
semaphore is greater than zero in any instant of execution of a
unit. In concurrent systems where synchronization is controlled by
balanced counter semaphores, like in behavior expressions here
defined, the value of a semaphore must be the same at the initial
and final states of the system.


\begin{figure}
\begin{center}
\begin{minipage}{\textwidth}
    \centering
    \screendump{1.0}{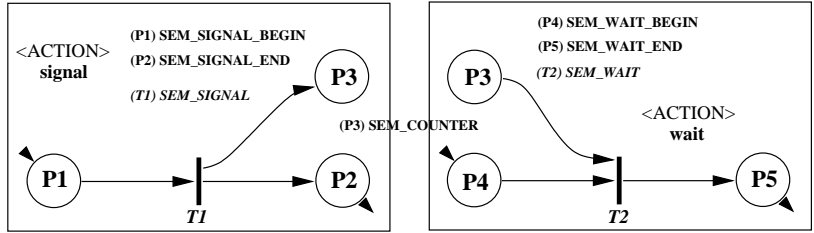}
\end{minipage}
\end{center}
\caption{\textbf{signal} (left) and \textbf{wait} (right) Semaphore
Primitives}

\label{fig:signal_translation_diagram}
\end{figure}

The \textbf{signal} and \textbf{wait} primitives are modelled
using Petri nets as described in Figure
\ref{fig:signal_translation_diagram}. Given a semaphore $s$, the
number of marks in place \textsc{sem\_counter}[s] models the
semaphore value at a given state.

\begin{figure}[b]
\begin{center}
\begin{minipage}{\textwidth}
\centering
    \screendump{0.6}{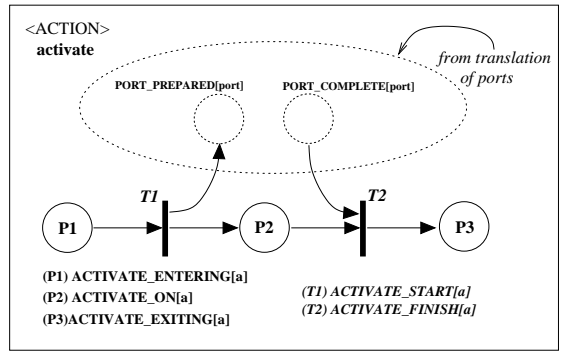}
\end{minipage}
\end{center}
\caption{Activation of Ports} \label{fig:activation_diagram}
\end{figure}

\subsubsection{Port Activation}
\label{sec:translation_port_activation}

The \textsf{Hash} primitive actions ! and ? models respectively \emph{send}
and \emph{receive} primitives in message passing programming. In
\textsf{Hash} programs, they cause activation of groups of ports and
individual ports that are not member of any group. In Section
\ref{sec:modelling_units}, the Petri net slice that model
individual ports and groups of ports is illustrated in Figure
\ref{fig:unit_ports_diagram}. An individual port $p$ is prepared
for communication whenever a mark is deposited in place
\textsc{port\_prepared}[$p$]. A group of ports of kind
\textbf{all} is prepared whenever all of its ports are prepared,
while a group of kind \textbf{any} is prepared whenever any of its
ports are prepared. The communication is completed whenever a
place is deposited in place \textsc{port\_complete}[$p$] or
\textsc{group\_complete}[$p$]. The activation $a$ of a port $p$ is
defined as the time between preparation of a port and completion
of communication. The format of Petri net slices induced by
translation of occurrences of primitives ! and ? is illustrated in
Figure \ref{fig:activation_diagram}. The firing of transition
\textsc{activate\_start}[$a$] prepares port $p$ for communication
and firing of transition \textsc{activate\_stop}[$a$] occurs
whenever the port completes communication. Notice that whenever a
mark is deposited in place \textsc{activate\_on}[$p$], the port
$p$ is active.

\begin{figure}
\begin{center}
\begin{minipage}{\textwidth}
\centering
    \screendump{0.57}{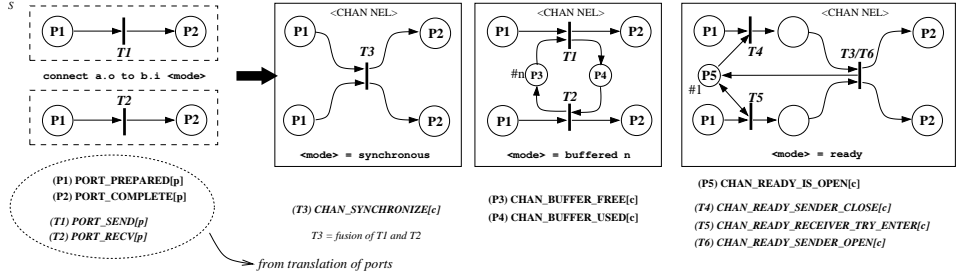}
\end{minipage}
\end{center}
\caption{Modelling Communication Channels}
\label{fig:channels_diagram}
\end{figure}

\subsection{Communication between Units}

In the translation of a component, the Petri net slices resulted
from translation of their units form an interlaced Petri net that
models their asynchronous execution. Information regarding
synchronization of units, by means of communication channels, is
not yet included. In Figure \ref{fig:channels_diagram}, Petri net
slices that model, respectively, the three kinds of channels that
may occur in a \textsf{Hash} program (\textbf{synchronous}, \textbf{buffered}
and \textbf{ready}) are presented. The translation function
$\Upsilon^L$ is applied to each communication channel in a
component, generating Petri net slices according to the
translation schema illustrated in Figure
\ref{fig:channels_diagram}. These Petri net slices are overlapped
with the Petri net slices that models behavior of units in order
to model synchronous execution of the units.


For \textbf{synchronous} channels, communication pairs must be
active at the same time for communication to complete. For
implementing that, it is only necessary to unify the respective
transitions modelling completion of communication from the
respective pairs. For \textbf{buffered} channels (bounded
buffers), the sender does not need to wait completion of
communication operation for resuming execution. For that, whenever
the sender port $s$ in a channel $c$ is activated, transition
\textsc{port\_send}[$s$] must be activated if there is a mark in
place \textsc{chan\_buffer\_free}[$c$], which models the number of
empty slots in the buffer. The place
\textsc{chan\_buffer\_used}[$c$], which receives a mark after
activation of \textsc{port\_send}[$s$], models the number of used
slots in the buffer. Notice that the sender blocks whenever there
is no empty slots in the buffer. Channels supporting
\textbf{ready} mode require a more complex protocol. The place
\textsc{chan\_ready\_is\_open}[$c$] ensures that communication
proceeds only if activation of the sender port precedes activation
of the receiver. Notice that whenever receiver is activated before
sender, the sender cannot proceed, causing a deadlock that may be
detected by a Petri net verification tool. Using this approach, it
is possible to verify, for example, if a certain parallel program
using \textbf{ready} channels may fail in some program state
during execution. Ready communication mode may improve
communication performance of MPI programs, but unfortunately it is
hard to ensure that communication semantics is safe in arbitrary
parallel programs. The modelling of communication semantics with
Petri nets may overcome these difficulties for debugging.

\begin{figure}
\begin{center}
\begin{minipage}{\textwidth}
\centering
    \screendump{0.35}{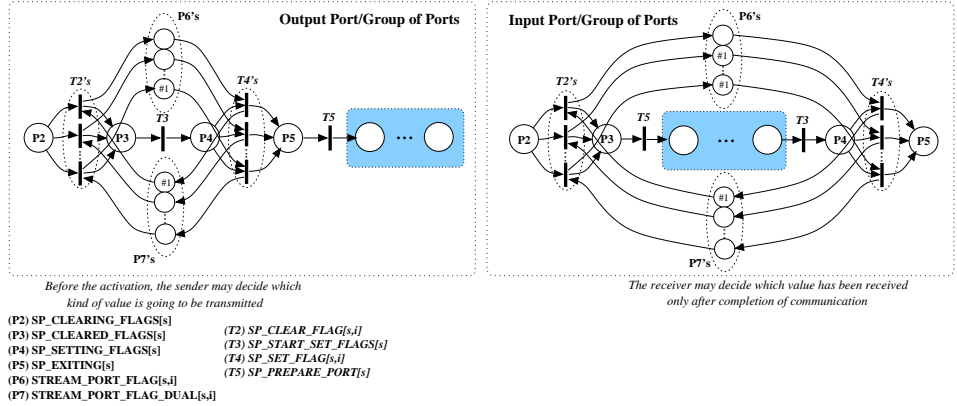}
\end{minipage}
\end{center}
\caption{Activation of Stream Ports}
\label{fig:activation_stream_port}
\end{figure}

\subsection{Synchronization of Streams Protocol}
\label{sec:stream_syncronization_protocol}

In Section \ref{sec:modelling_stream_communication}, it was
introduced two sets of places that must exist for each stream port
$s$: ${\textsc{Stream\_Flags}}(s)$ and
$\overline{\textsc{Stream\_Flags}}(s)$. Additionally, restrictions
were introduced for their markings. They allow to check the kind
of the transmitted value in the last activation of the port,
making possible to check stream termination conditions that occurs
in \textbf{repeat} and \textbf{if} combinators. This section
presents a protocol for updating marking of places
${\textsc{Stream\_Flags}}(s)$ and
$\overline{\textsc{Stream\_Flags}}(s)$ in such a way that
restrictions introduced in Section
\ref{sec:modelling_stream_communication} are obeyed.

In Figure \ref{fig:activation_stream_port}, it is illustrated how
the network presented in Figure \ref{fig:activation_diagram}
(activation of ports) may be enriched in order to introduce a
protocol for updating places in ${\textsc{Stream\_Flags}}(s)$ and
$\overline{\textsc{Stream\_Flags}}(s)$, for an arbitrary stream
port $s$ with nesting factor $n$. The Petri net slice introduced
have the transitions \textsc{sp\_clear\_flag}[$s$,$i$] e
\textsc{sp\_set\_flag}[$s$,$i$], for $0 \leq i \leq n$ as its main
components. They are arranged in such a way that transitions
inside the sets are mutually exclusive. The firing of a transition
in \textsc{sp\_clear\_flag}[$s$,$i$] clears the set of places
${\textsc{Stream\_Flags}}(s)$, by moving the mark in the
corresponding cleared place to the corresponding place in set
$\overline{\textsc{Stream\_Flags}}(s)$. After that, all places in
$\overline{\textsc{Stream\_Flags}}(s)$ have exactly one mark,
while all places in ${\textsc{Stream\_Flags}}(s)$ have zero mark.
In sequence, one transition of set \textsc{sp\_set\_flag}[$s$,$i$]
is fired causing the moving of a mark from one of the places in
$\overline{\textsc{Stream\_Flags}}(s)$, chosen
non-deterministically, to the corresponding place in
${\textsc{Stream\_Flags}}(s)$. This sequence of actions models the
test of the kind of the value transmitted in the current
activation of the port. Notice that the moment of the choice is
different for input and output ports. An input port only updates
its ${\textsc{Stream\_Flags}}(s)$ after communication has been
completed. This is in accordance with implementation semantics,
once only after receiving the value, the receiver may check the
kind of the transmitted value.

\begin{figure}
\begin{center}
\begin{minipage}{\textwidth}
\centering
    \screendump{0.38}{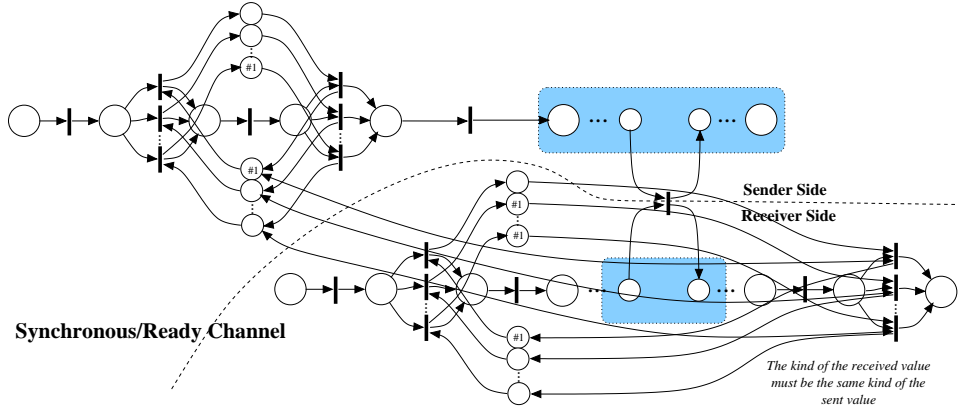}
\end{minipage}
\end{center}
\caption{Ensuring the Consistency of Streams Connected through
Synchronous Channels} \label{fig:stream_channel_consistency_sync}
\end{figure}

\subsubsection{Ensuring Consistency of Communication}

Channel communication semantics imposes that the kind of value
transmitted by the sender in a given activation of a stream port
is the same as the kind of the value that the receiver receives in
the corresponding activation.

In Figure \ref{fig:stream_channel_consistency_sync}, it is shown
how this restriction is ensured for channels with
\textbf{synchronous} and \textbf{ready} mode of communication.

For the following description, consider that individual ports are
groups containing only one port. Thus, consider a channel $c$
connecting a sender stream port $s$ and a receiver stream port
$r$, both with nesting factor $n$. The groups where $s$ and $r$
are contained are respectively $g_s$ and $g_r$.

For each $i$, $0 \leq i \leq n$, it is necessary to create an arc
that links the place \textsc{stream\_port\_flag}[$g_s$,$i$] to the
transition \textsc{sp\_set\_flag}[$g_r$,$i$], in such way that
consistency of ${\textsc{Stream\_Flags}}(s)$ and
${\textsc{Stream\_Flags}}(r)$ is forced after communication
completion. If two different senders, connected to ports belonging
to a group $g_r$ of kind \textbf{all}, decide to transmit values
of different nesting levels in a give activation of $g_r$, a
deadlock occurs. But it was possible to introduce a Petri net
slice for detecting this event.

\begin{figure}
\begin{center}
\begin{minipage}{\textwidth}
\centering
    \screendump{0.38}{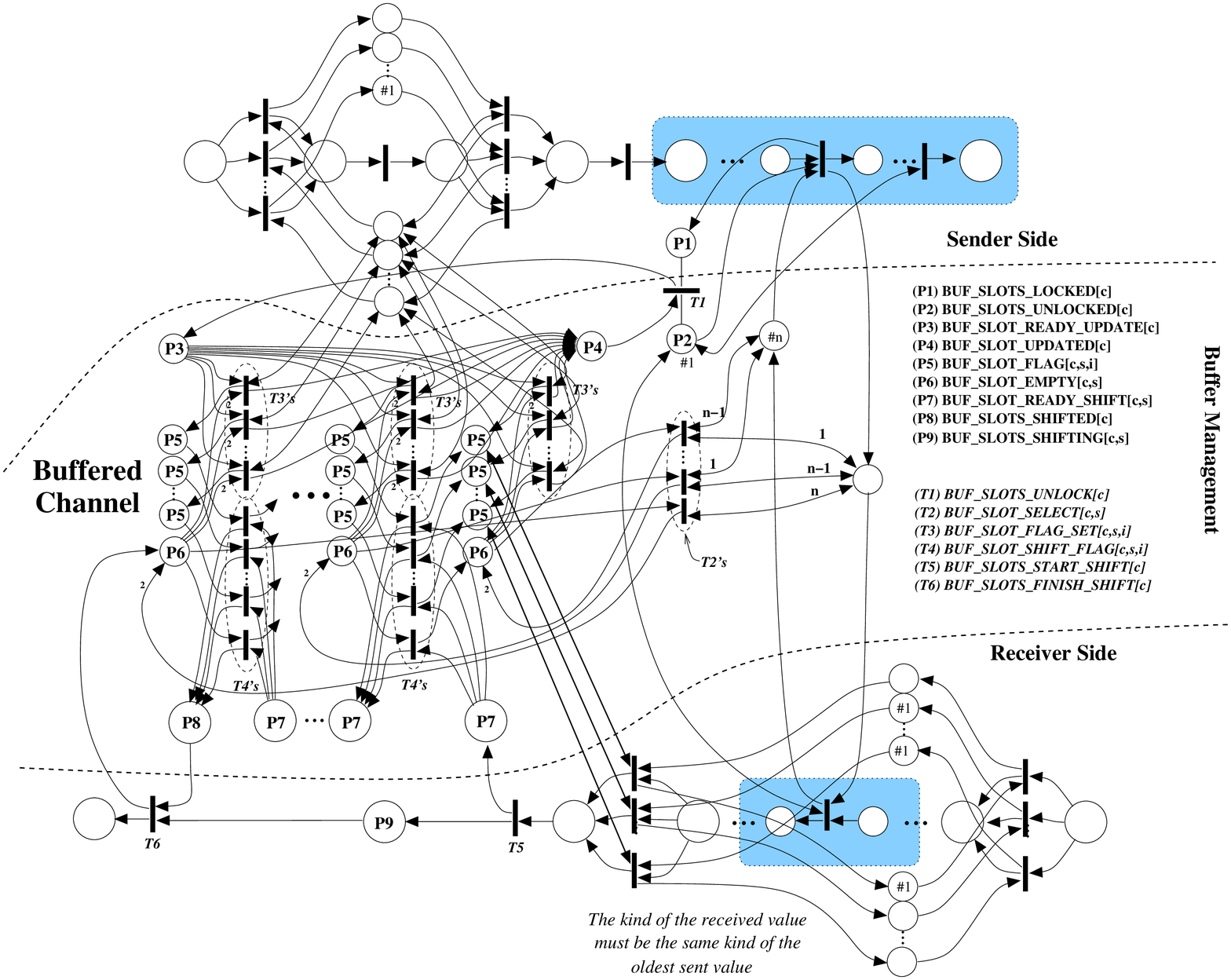}
\end{minipage}
\end{center}
\caption{Ensuring Consistency Streams Connected through Buffered Channels} \label{fig:stream_channel_buf}
\end{figure}

Bounded buffered communication imposes a more complicated
approach, illustrated in Figure \ref{fig:stream_channel_buf}.
Consider that channel $c$ have a buffer of size $b$. For each
buffer slot $k$, $0 \leq k \leq b-1$, in a channel connecting
ports with nesting factor $n$, there is a set of places
\textsc{buf\_slot\_flag}[$c$,$k$,$i$], for $0 \leq i \leq n$. They
remember the kind of value stored in a buffer slot $k$, after
communication in a channel $c$. Essentially, after firing
transition \textsc{sp\_set\_flag}[$g_s$,$i$], the marking of
\textsc{stream\_port\_flag}[$g_s$,$i$] at the moment of the
activation of the group $g_s$ is saved in
\textsc{buf\_slot\_flag}[$c$,$k$,$i$], where $k$ is the number of
the next available buffer slot. If all slots are filled, the
sender blocks until a slot is freed by the receiver.

\begin{figure}
\begin{center}
\begin{minipage}{\textwidth}
    \centering
    \screendump{0.45}{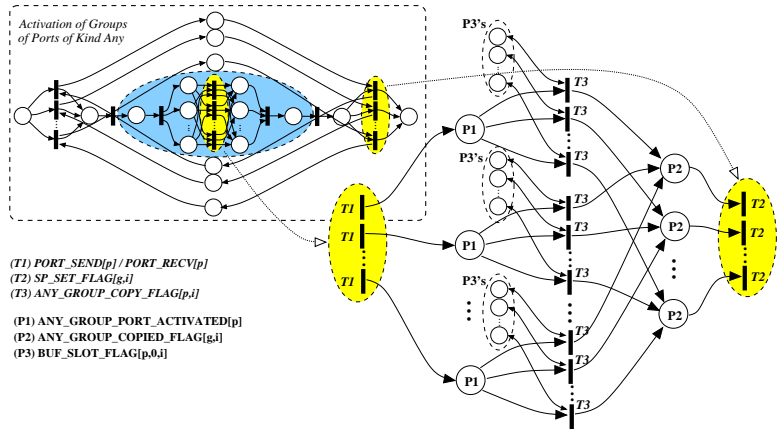}
\end{minipage}
\end{center}
\caption{Copying Protocol for Groups of Input Ports of Kind \textbf{any}}
\label{fig:any_groups_protocol_diagram}
\end{figure}

If $g_r$ is a group of kind \textbf{all} (notice that all groups
with one port is of kind \textbf{all}), an arc from each place
\textsc{buf\_slot\_flag}[$c$,0,$i$] to transition
\textsc{sp\_set\_flag}[$g_r$,$i$] ensures that the marking of
${\textsc{Stream\_Flags}}(r)$ reflects the kind of the oldest
value placed in the buffer by the sender, as semantics of buffered
communication imposes. If $g_r$ is a group of kind \textbf{any},
it is necessary to introduce the Petri net slice shown in Figure
\ref{fig:any_groups_protocol_diagram}. It ensures to copy the
marking of the places \textsc{buf\_slot\_flag}[$c$,0,$i$] of the
chosen receiver port $r$ in $g_r$. For that, the mutually
exclusive places \textsc{any\_group\_port\_activated}[$r$], for
each $r$ belonging to $g_r$, remember which port of $g_r$ was
chosen. They enable the appropriate set of transitions
\textsc{any\_group\_copy\_flag}[$r$], which are connected to
places \textsc{buf\_slot\_flag}[$c$,$0$,$i$] of the channel where
the chosen port is connected. The marking of this set of places is
copied to places \textsc{any\_group\_copied\_flag}[$r$,$i$], which
are connected to transitions \textsc{sp\_set\_flag}[$g_r$, $i$].

The arrangement of the places \textsc{buf\_slots\_locked}[$c$] and
\textsc{buf\_slots\_unlocked}[$c$] and the transition
\textsc{buf\_slots\_unlock}[$c$] avoids accesses to the buffer
while it is being updated after a transmission. The firing of a
transition \textsc{buf\_slot\_select}[$c$,$k$] means the selection
of the next empty slot. Whenever the buffer is full ($n$ marks are
deposited in place \textsc{chan\_buffer\_used}[$c$]), the sender
must block for waiting that the receiver consume the contents of
the first slot entry.

After copying the kind of the value in the first slot of the
buffer, it is discarded. For that, a shift operation, that occurs
while a mark is deposited in place
\textsc{buf\_slots\_shifting}[$c$], allows to save the marking of
places \textsc{buf\_slot\_flag}[$c$,$k$,$i$] to
\textsc{buf\_slot\_flag}[$c$,$k-1$,$i$], for $1 \leq k \leq n$.

\subsubsection{Ensuring Consistency of Order of Kind of Transmitted Values}

The following example illustrates the needs for imposing one more
restriction in the Petri net slice of the stream control protocol.
For instance, consider a nested Haskell list of \texttt{Int}'s
with nesting factor 4.

\begin{center}
\begin{footnotesize}
\begin{tt}
\begin{minipage} {\textwidth}
\begin{tabbing}

[[[[1],[5,6]],[[2,3]]],[],[[[[4,5,7],[8,9]]],[[6],[7,9]]]]

\end{tabbing}
\end{minipage}
\end{tt}
\end{footnotesize}
\end{center}

The correspondent stream in the \textsf{Hash} component model will transmit the
following values at each activation of the corresponding port:

\begin{center}
\begin{footnotesize}
\begin{tt}
\begin{minipage} {\textwidth}
\begin{tabbing}

\{\=1,\textsc{Eos} 3, 5, 6,\textsc{Eos} 3,\textsc{Eos} 1, 2, 3,\textsc{Eos} 3,\textsc{Eos} 2,\textsc{Eos} 1,\textsc{Eos} 1, 4, 5, 7, \\
  \>\textsc{Eos} 3, 8, 9,\textsc{Eos} 3,\textsc{Eos} 2,\textsc{Eos} 1, 6,\textsc{Eos} 3, 7, 9,\textsc{Eos} 3,\textsc{Eos} 2,\textsc{Eos} 1,\textsc{Eos} 0\} \\

\end{tabbing}
\end{minipage}
\end{tt}
\end{footnotesize}
\end{center}

Notice that after transmitting a value \textsc{Eos $2$}, it is not
possible to transmit a value \textsc{Eos $0$}, since the enclosing
stream at nesting level $1$ was not finalized yet. In this case,
only values \textsc{Eos $1$}, \textsc{Eos $2$}, \textsc{Eos $3$}
and a data item may be transmitted.

Now, consider the general case for the transmission of given end
marker at nesting level $k$. If $k \geq 1$, the next value to be
transmitted may be any of the end markers at nesting level greater
than $k-1$ or a data item. If $k = 0$, the stream is
\emph{finalized}. Any attempt to read a value in a finalized
stream is considered an error.

\begin{figure}
\begin{center}
\begin{minipage}{\textwidth}
\centering
    \screendump{0.575}{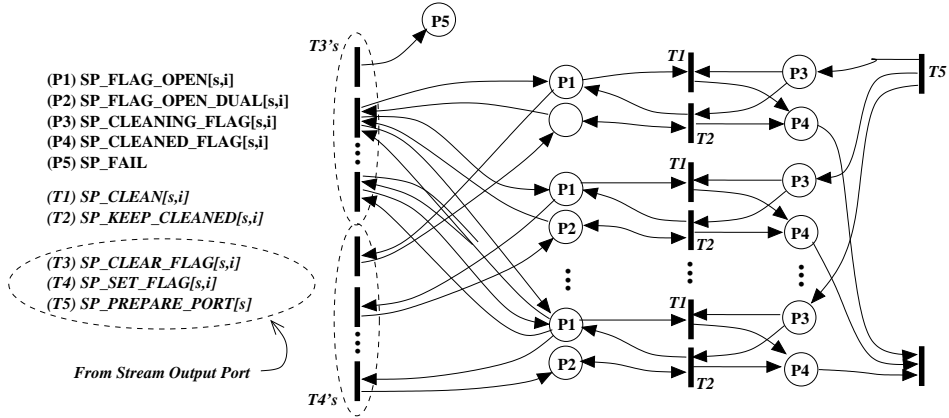}
\end{minipage}
\end{center}
\caption{Consistency of the Order of Kind of Transmitted Stream
Values} \label{fig:stream_consistency_order}
\end{figure}

In Figure \ref{fig:stream_consistency_order}, it is shown how the
Petri net slice presented in Figure
\ref{fig:activation_stream_port} (output port) may be enriched in
order to support the restriction stated in the last paragraph. A
mark is placed in place \textsc{sp\_order\_fail}[$s$] whenever an
attempt to activate a finalized stream port $s$ (the kind of last
transmitted value is \textsc{Eos $0$}) occurs. For each nesting
level $i$, four places and two transitions controls the
consistency of the order of the transmitted value. The place
\textsc{sp\_flag\_open}[$s$,$i$] receives a value whenever a value
\textsc{Eos $i$} may be sent. Its mutually exclusive dual place,
named \textsc{sp\_flag\_open\_dual}[$s$,$i$] allows for resetting
the marking of \textsc{sp\_flag\_open}[$s$,$i$] after activation.
Resetting procedure is implemented using the next elements
described here. The place \textsc{sp\_cleaning\_flag}[$s$,$i$] has
a mark whenever the resetting procedure is enabled for the port
$s$ in nesting level $i$. Its corresponding place
\textsc{sp\_cleaned\_flag}[$s$,$i$] has a mark after resetting
procedure finishes. The transition \textsc{sp\_clean}[$s$,$i$]
resets the place \textsc{sp\_flag\_open}[$s$,$i$] to its original
state (zero marks), while transition
\textsc{sp\_keep\_cleaned}[$s$,$i$] fires whenever the place
\textsc{sp\_flag\_open}[$s$,$i$] is already cleaned. Notice that
the transitions \textsc{sp\_keep\_cleaned}[$s$,$i$] and
\textsc{sp\_clean}[$s$,$i$] are mutual exclusive, since
\textsc{sp\_flag\_open}[$s$,$i$] and
\textsc{sp\_flag\_open\_dual}[$s$,$i$] are too.


\subsection{The Complexity of the Generated Petri Net}

After overlapping the Petri net slice that models stream
communication semantics, allowing to make precise analysis about
behavior of \textsf{Hash} programs at coordination level, the Petri net of
simple \textsf{Hash} programs may become very large. Large Petri nets may
turn impossible for programmers the analysis without help of some
automatic or higher level means. Also, it makes hard and memory
consuming the computations performed by the underlying Petri net
tools, such as computation of reachability and coverability
graphs, place and transition invariants, etc. These difficulties
comes from transitory technological limitations, since processing
power and memory amount of machines have increased rapidly in the
recent years and it is expected that they will continue to
increase in the next decades. Also, it is possible to use parallel
techniques to perform high computing demanding analysis, but this
approach has been exploited by few designers of Petri net tools.
It is a reasonable assumption that parallel programmers have
access to some parallel computer.

Despite these facts, it is desirable to provide ways for helping
programmers to work with large Petri nets or simplifying the
generated Petri net. The following techniques have been proposed:

\begin{itemize}

\item During the process of analysis, the programmer may decide
not to augment the Petri net of the \textsf{Hash} program with the protocol
for modelling stream communication. This approach makes sense
whenever the information provided by the protocol is not always
necessary for the analysis being conducted. This is the reason for
the separation of the stream communication protocol from the rest
of the Petri net in the \textsf{Hash} program;

\item Another approach that have been proposed, but for further
works, is to build higher level environments for analysis of \textsf{Hash}
programs on top of Petri net tools. Instead of programmers to
manipulate Petri net components, they manipulate \textsf{Hash} program
elements of abstraction. Then, the analysis are transparently and
automatically translated into proving sequences using INA tool;

\item Specific translation schemas for skeletons might be
specified in order to simplify the generated Petri net. This
approach is illustrated in the next section. For instance, it is
defined how higher-level information provided by the use of
collective communication skeletons might be used in the
translation process of \textsf{Hash} programs into Petri nets.

\end{itemize}

\begin{figure}

\begin{center}
\begin{tiny}
\begin{minipage} {\textwidth}
\begin{tabbing}

\texttt{1.}\ \ \ \= \textbf{component} IS$<$\= \textsc{problem\_class, num\_procs,} \= \textsc{max\_key\_log2, num\_buckets\_log2}, \\
\texttt{2.} \>                        \>                                     \> \textsc{total\_keys\_log2, max\_iterations, max\_procs, test\_array\_size}$>$ \textbf{with}\\
\texttt{3.} \> \\
\texttt{4.} \> \#define PARAMETERS (\texttt{IS\_Params} \=\textsc{problem\_class} \textsc{num\_procs} \= \textsc{max\_key\_log2} \textsc{num\_buckets\_log2} \\
\texttt{5.} \>                                         \>\>\textsc{total\_keys\_log2} \textsc{max\_iterations} \textsc{max\_procs} \textsc{test\_array\_size}) \\
\texttt{6.} \> \\
\texttt{7.} \> \textbf{iterator} i \textbf{range} [1, \textsc{num\_procs}] \\
\texttt{8.} \> \\
\texttt{9.} \> \textbf{use} \textsc{Skeletons.\{Misc.RShift, Collective.\{AllReduce, AllToAllv\}\}} \\
\texttt{10.}\>  \textbf{use} IS\_FM         \textit{$--$ IS Functional Module} \\
\texttt{11.}\>  \\
\texttt{12.}\>  \textbf{unit} bs\_comm \= \textbf{assign} \textsc{AllReduce}$<$\textsc{num\_procs}, \texttt{MPI\_SUM}, \texttt{MPI\_INTEGER}$>$ \= \textbf{to} bs\_comm \\
\texttt{13.}\>  \textbf{unit} kb\_comm \> \textbf{assign} \textsc{AllToAllv}$<$\textsc{num\_procs}$>$                                           \> \textbf{to} kb\_comm \\
\texttt{14.}\>  \textbf{unit} k\_shift \> \textbf{assign} \textsc{RShift}$<$\textsc{num\_procs}$>$ 0 $\rightarrow$ \_                           \> \textbf{to} k\_shift \\
\texttt{15.}\>  \\
\texttt{16.}\>  \textbf{interface} \= \emph{IIS} (bs*, kb*, k) $\rightarrow$ (bs*, kb*, k) \\
\texttt{17.}\>                    \> \textbf{where}: bs@\emph{IAllReduce} (UArray Int Int) \# kb@\emph{IAllToAllv} (Int, Ptr Int) \# k@\emph{RShift} Int \\
\texttt{18.}\>                    \> \textbf{behaviour}: \textbf{seq} \{\=\textbf{repeat} \textbf{seq} \{\textbf{do} bs; \textbf{do} kb\} \textbf{until} $<$bs \& kb$>$; \textbf{do} k\} \\
\texttt{19.}\>  \\
\texttt{20.}\>  $[/$ \= \textbf{unify} \= bs\_comm.p[i] \# bs, kb\_comm.p[i] \# kb, k\_comm.p[i] \# k \textbf{to} is\_peer[i] \= \# \emph{IIS} \\
\texttt{21.}\>       \> \textbf{assign} IS\_FM (PARAMETERS, bs, kb, k) $\rightarrow$ (bs, kb, k) \textbf{to} is\_peer[i] \# bs \# kb \# k $/]$ 

\end{tabbing}
\end{minipage}
\end{tiny}
\end{center}

\caption{The Configuration Code of IS Program} \label{fig:is_code}

\end{figure}

\subsection{Modelling Collective Communication Skeletons}

Message passing libraries, such as MPI, support special primitives
for collective communication. In \textsf{Hash} programming environment, it is
defined a library of skeletons that implement the pattern of
communication involved in collective communication operations
supported by MPI. They are: \textsc{Bcast}, \textsc{Scatter},
\textsc{Scatterv}, \textsc{Reduce\_Scatter}, \textsc{Scan},
\textsc{Gather}, \textsc{Gatherv}, \textsc{Reduce},
\textsc{AllGather}, \textsc{AllGatherv}, \textsc{AllToAll},
\textsc{AllToAllv}, and \textsc{AllReduce}. In the next
paragraphs, the use of collective communication skeletons is
illustrated by means of an example.

In Figure \ref{fig:is_code}, the code for the \textsf{Hash} version of IS
program, from NPB (NAS Parallel Benchmarks) \cite{Bailey1995} is
presented. IS is a parallel implementation of the bucket sort
algorithm, originally written in C/MPI. In this section, IS is
used to illustrate the use of collective communication skeletons
in a \textsf{Hash} program, motivating definition of a specific strategy for
translating collective communication patterns of interaction among
units in a \textsf{Hash} program.

\begin{figure}
\begin{center}
\begin{minipage}{\textwidth}
\centering
    \screendump{0.575}{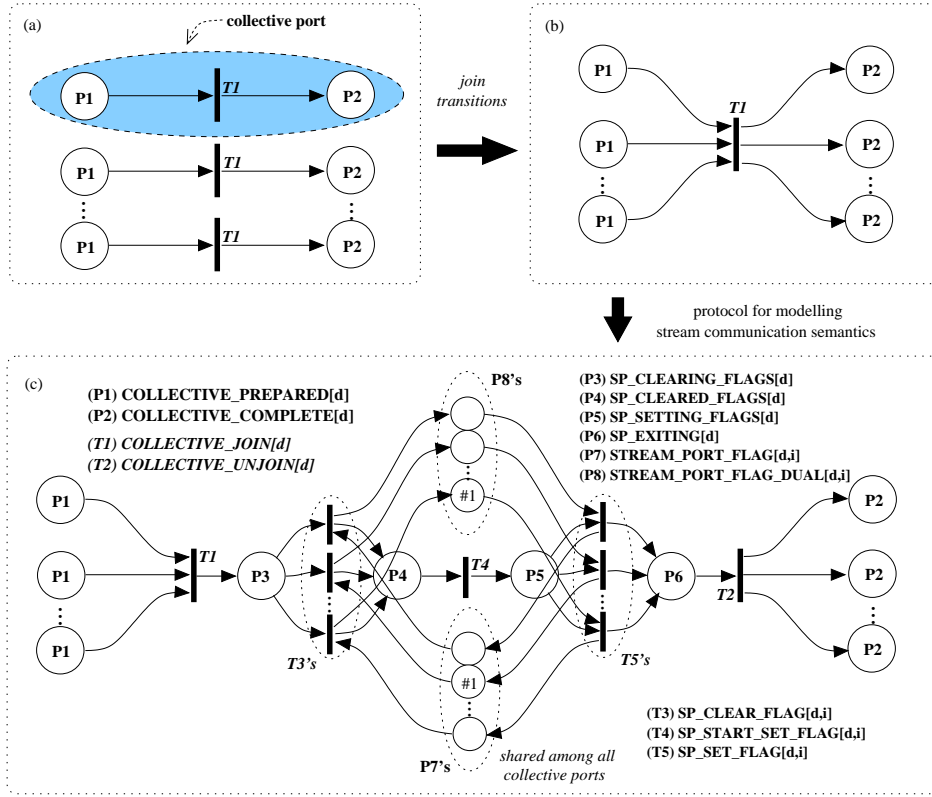}
\end{minipage}
\end{center}
\caption{Modelling Collective Communication Skeleton Semantics}
\label{fig:skeleton_translation}
\end{figure}

In the line 9 of Figure \ref{fig:is_code}, it is declared that
skeletons\footnote{Partial topological skeletons are composed
components where at least one unit is virtual. In the case of
collective communication skeletons, all units are virtual.}
\textsc{AllReduce} and \textsc{AllToAllv} will be used in the
configuration. In lines 12, 13, and 14, three units are declared,
named \emph{bs\_comm}, \emph{kb\_comm} and \emph{k\_shift}. The
first two have collective communication skeletons assigned to
them, respectively \textsc{AllReduce} and \textsc{AllToAllv}.
Since skeletons are composed components, these units are
\emph{clusters} of units that interact using the collective
communication patterns described by the skeleton. In line 20, the
unification of correspondent units that comprise clusters
\emph{bs\_comm}, \emph{kb\_comm} and \emph{k\_shift} forms the
units that comprise the IS topology. The unification of virtual
units from distinct cluster allows to overlap skeletons. The
behavior of the virtual units that result from unification, named
\emph{is\_peer[i]}, $1 \leq i \leq \textsc{num\_procs}$, is
specified by interface \emph{IIS}. The simple component
\textsc{IS\_FM} is assigned to them for defining the computation
of each process in the IS program. Notice that IS is a SPMD
program, where the task performed by all processes is defined by
the same simple component \cite{Carvalho2003f}.

The interface \emph{IIS} is declared from composition of
\emph{IAllReduce}, \emph{IAllToAllv} and \emph{IRShift}
interfaces, the interface slices of \emph{IIS}. The interface
slices are respectively identified by \emph{bs}, \emph{kb}, and
\emph{k}. The use of combinator \textbf{do} is an abbreviation
that avoids to rewrite the behavior of interface slices. Thus,
``\textbf{do} \emph{bs}'' relates to the sequence of actions
encapsulated in the specification of interface \emph{IAllReduce}.

Using the above conventions for overlapping collective
communication skeletons in order to form more complex topologies,
it is simple to define a specific translation rule for patterns of
collective communication interactions. The identifiers of
interface slices from collective communication skeletons might be
viewed as special kinds of ports (collective ports) and the
operator \textbf{do} as its activation operator. Notice that
interface slices may be used in termination conditions of streams,
like in line 18. Collective ports have no direction, since all
processes participate in communication. All communication
operations are synchronous. The Figure
\ref{fig:skeleton_translation}(b) illustrates a Petri net slice
that models a collective communication operation. The involved
ports are \emph{collective ports} that correspond to interface
slices of units that participates in a certain collective
communication operation, defined by the cluster of units that
defines it. In Figure \ref{fig:skeleton_translation}(c), it is
illustrated how to augment the Petri net slice shown in Figure
\ref{fig:skeleton_translation}(b) with a protocol for modelling
stream communication semantics. It is important to notice that the
set of places $\textsc{Stream\_Flags}(s)$ and
$\overline{\textsc{Stream\_Flags}}(s)$ are shared by all
collective ports involved in a collective operation.

\section{Petri Net Analysis of Formal Properties}
\label{sec:formal_properties_analysis}

In this section, solutions for two well known synchronization
problems, implemented in \textsf{Hash} approach, illustrates the use of Petri
nets for analyzing \textsf{Hash} programs by verifying their formal
properties.

\subsection{Dining Philosophers}

The \emph{dining philosophers problem} is one of the most relevant
synchronization problems in concurrency theory. Since it was
originally proposed by Dijkstra in 1968 \cite{Dijkstra1968}, it
has been widely used for exercising the ability of concurrent
languages and models for providing elegant solutions for avoiding
deadlocks in concurrent programs. The dining philosophers problem
is stated in the following way:

\begin{quote}

Five philosophers are sited around a table for dinner. The
philosophers spend their times \emph{eating} and \emph{thinking}.
When a philosopher wants to eat, he takes two forks from the
table. When a philosopher wants to think, he keeps the two forks
available on the table. However, there are only five forks
available, requiring that each philosopher share their two forks
with their neighbors. Thus, whenever a philosopher is eating, its
neighbors are thinking.

\end{quote}

\begin{figure}
\begin{center}

\begin{tabular}{cc}
\screendump{0.6}{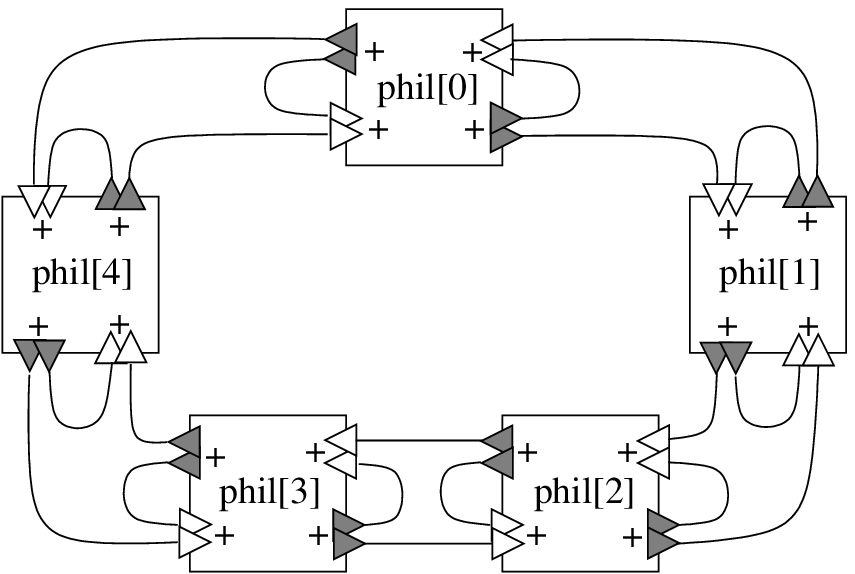} & \screendump{0.6}{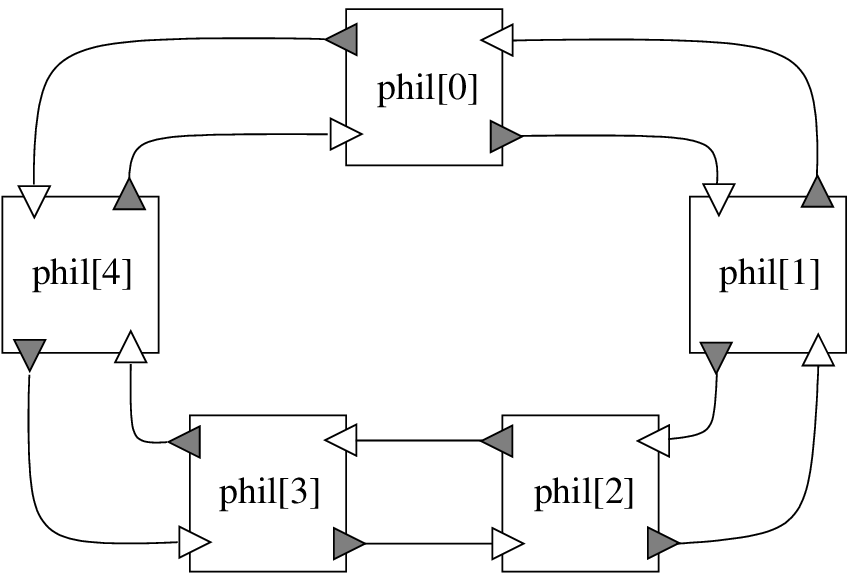} \\
(a) & (b)
\end{tabular}

\end{center}

\caption{\textsf{Hash} Topologies for the Dining Philosophers Problem}

\label{fig:dining_philosophers_topology}
\end{figure}

\begin{figure}
\begin{center}
\begin{footnotesize}

\begin{tabular}{c}

\hline

\begin{minipage} {\textwidth}
\begin{tabbing}
\\
\textbf{component} DiningPhilosophers$<$\texttt{N}$>$ \textbf{with} \\
\\
\textbf{index} i \textbf{range} [0,\texttt{N}-1] \\
\\
\textbf{interface} \emph{IPhil} \textbf{where} \= \textbf{ports}: (rf\_get*, lf\_get*) $\rightarrow$ (rf\_put*, rf\_put*)\\
                   \> \textbf{protocol}: \= \textbf{seq} \{\=rf\_put!; \\
                   \>                       \>                \>\textbf{repeat} \textbf{seq} \{\=\textbf{par} \{lf\_get?; rf\_get?\};  \\
                   \>                       \>                \>                               \>\textbf{par} \{lf\_put!; rf\_put!\}\} \\
                   \>                       \>                \> \textbf{until} $<$ lf\_get \& rf\_get \& lf\_put \& rf\_put $>$\} \\
\\
$[/$ \textbf{unit} phil[i] \textbf{where} \= \textbf{ports}: \emph{IPhil} \\
                                          \> \textbf{grouping}: \= rf\_get \{neighbor,self\} \textbf{any}, lf\_get \{neighbor,self\} \textbf{any}, \\
                                          \>                    \> rf\_put \{neighbor,self\} \textbf{any}, lf\_put \{neighbor,self\} \textbf{any} $/]$ \\
\\
$[/$ \= \textbf{connect} phil[i]$\rightarrow$rf\_put[neighbor] \= \textbf{to} phil[i-1 \textbf{mod} \texttt{N}]\ \ \= $\leftarrow$lf\_get[neighbor] \=, \textbf{buffered} 1 \\
     \> \textbf{connect} phil[i]$\rightarrow$rf\_put[self]     \> \textbf{to} phil[i]                              \> $\leftarrow$rf\_get[self]     \>, \textbf{buffered} 1 \\
     \> \textbf{connect} phil[i]$\rightarrow$lf\_put[neighbor] \> \textbf{to} phil[i+1 \textbf{mod} \texttt{N}]    \> $\leftarrow$rf\_get[neighbor] \>, \textbf{buffered} 1 \\
     \> \textbf{connect} phil[i]$\rightarrow$lf\_put[self]     \> \textbf{to} phil[i]                              \> $\leftarrow$lf\_get[self]     \>, \textbf{buffered} 1 $/]$ 

\end{tabbing}
\end{minipage}
\\
\\
(a)\\
\\
\hline

\begin{minipage} {\textwidth}
\begin{tabbing}
\\
\textbf{component} DiningPhilosophers$<$\texttt{N}$>$ \textbf{with} \\
\\
\textbf{index} i \textbf{range} [0,\texttt{N}-1] \\
\\
\textbf{interface} \emph{IPhil}[0] \textbf{where} \= \textbf{ports}: (rf\_get, lf\_get) $\rightarrow$ (rf\_put, rf\_put)\\
                   ~                              \> \textbf{protocol}: \textbf{repeat} \textbf{seq} \{lf\_put!;rf\_get?; lf\_get?; rf\_put!\} \\
\\
\textbf{interface} \emph{IPhil}[1] \textbf{where} \= \textbf{ports}: (rf\_get, lf\_get) $\rightarrow$ (rf\_put, rf\_put)\\
                                                  \> \textbf{protocol}: \textbf{repeat} \textbf{seq} \{rf\_get?; lf\_put!; rf\_put!; lf\_get?\} \\
\\
$[/$ \textbf{unit} phil[i] \textbf{where} \textbf{ports}: \emph{IPhil}[i \textbf{mod} 2] $/]$ \\
\\
$[/$ \= \textbf{connect} phil[i]$\rightarrow$rf\_put \= \textbf{to} phil[i-1 \textbf{mod} \texttt{N}]$\leftarrow$lf\_get, \textbf{buffered} \\
     \> \textbf{connect} phil[i]$\rightarrow$lf\_put \> \textbf{to} phil[i+1 \textbf{mod} \texttt{N}]$\leftarrow$rf\_get, \textbf{buffered} $/]$\\

\end{tabbing}
\end{minipage}
\\
\\
(b)
\\
\\
\hline

\end{tabular}

\end{footnotesize}
\caption{\textsf{Hash} Code for the First Solution of the Dining Philosophers
Problem} \label{fig:dining_philosophers_code_1}
\end{center}
\end{figure}

A solution to the dining philosopher problem establishes a
protocol for ordering the activity of the philosophers. In Figure
\ref{fig:dining_philosophers_topology}, \textsf{Hash} topologies for
solutions for the dining philosophers problem are presented. The
first one, whose code is presented in Figure
\ref{fig:dining_philosophers_code_1}(a), is an ``anarchical''
solution, where philosophers are free to decide when to think or
to eat. In this solution, the reader may observe the use of
buffered channels and groups of ports of kind \textbf{any}
composing the network topology. The one-slot buffered channel
allows to model the fact that a fork may be on the table waiting
for a philosopher to acquire them. This occurs whenever there is
one message pending on the buffer. At the beginning of the
interaction, all philosophers have a fork and release them. When a
philosopher releases a fork, the semantics of group of ports of
kind \textbf{any} ensures that if there is a philosopher waiting
for the fork, he will obtain the fork immediately. Otherwise, it
is possible that the philosopher that released the fork have a
chance to acquire the fork again. This solution does not satisfy
some of the enunciated requisites for a good solution to an
instance of the critical section problem. For example, if all
philosophers acquire their right (left) forks, they will be in
deadlock. It is also possible that a philosopher never get a
chance to obtain the forks (eventual entry). The second solution,
whose code is presented in Figure
\ref{fig:dining_philosophers_code_1}(b), ensure all requisites.
Additionally, it ensures maximal parallelism. In any state of
execution, there are two philosophers eating.

\begin{figure}
\begin{center}

\begin{tabular}{c}
    \screendump{0.25}{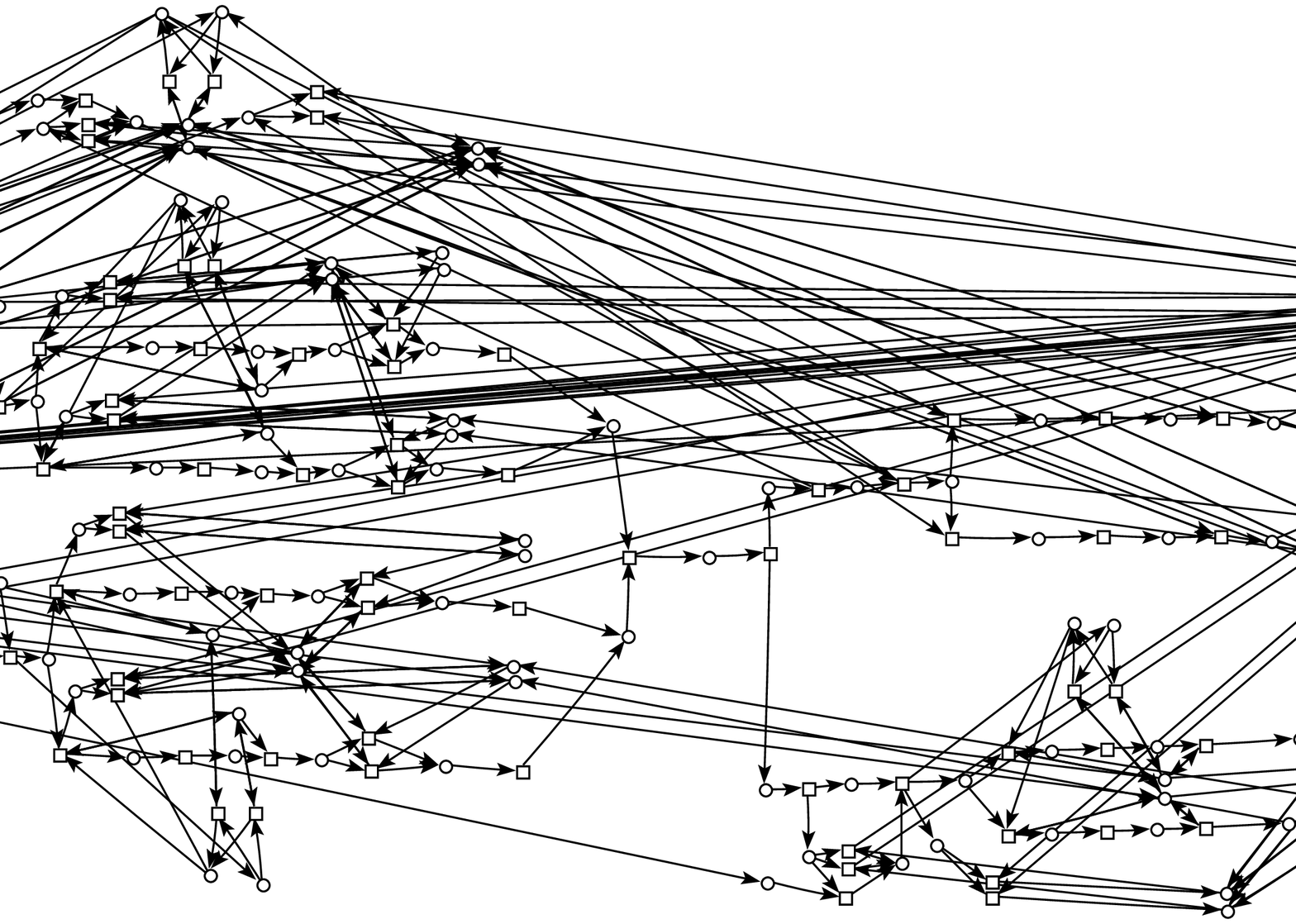} \\
    (a) \\
    \\
    \screendump{0.40}{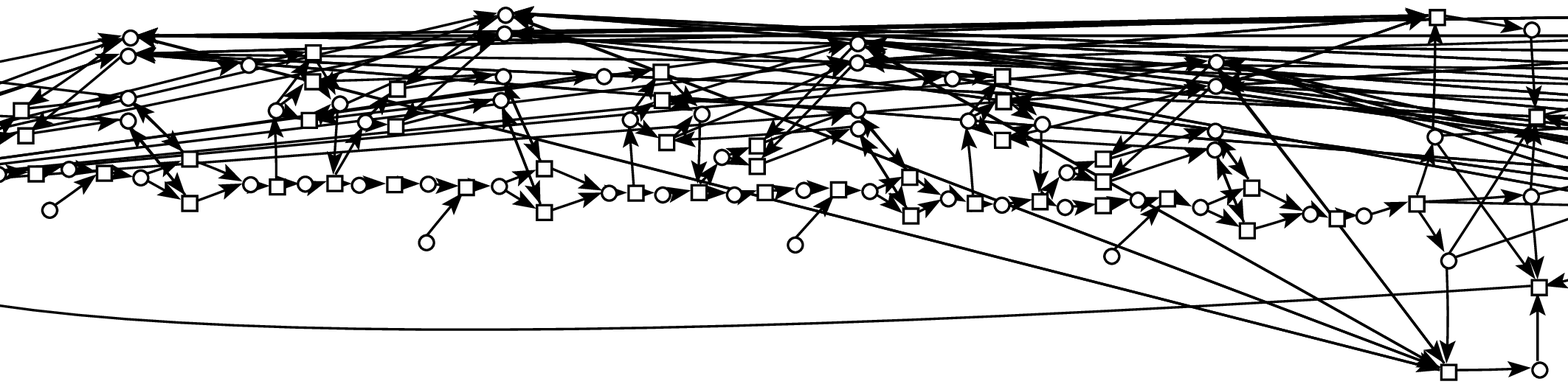} \\
    (b)
\end{tabular}

\end{center}

\caption{Petri Net Modelling Behavior of One Philosopher}

\label{fig:one_phil_translation}
\end{figure}

\begin{figure}
\begin{center}

\begin{tabular}{c}
    \screendump{0.20}{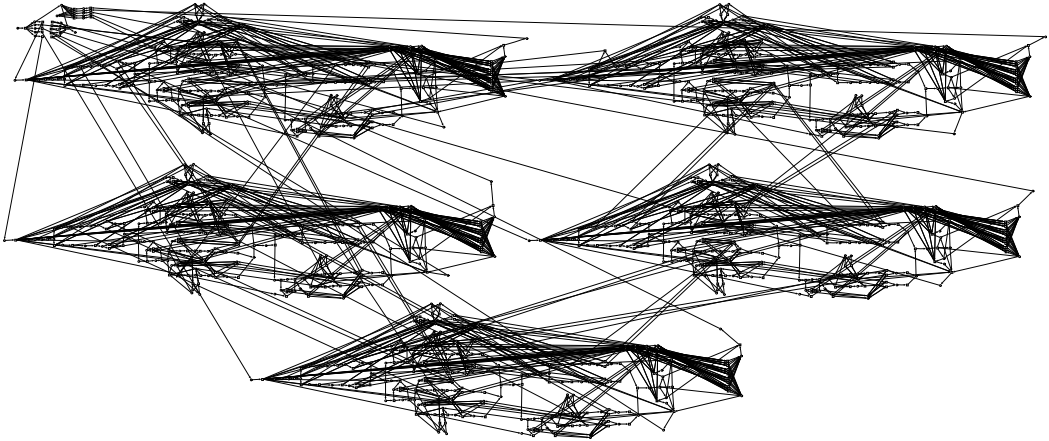} \\ \\
    (a) \\
    \\
    \screendump{0.40}{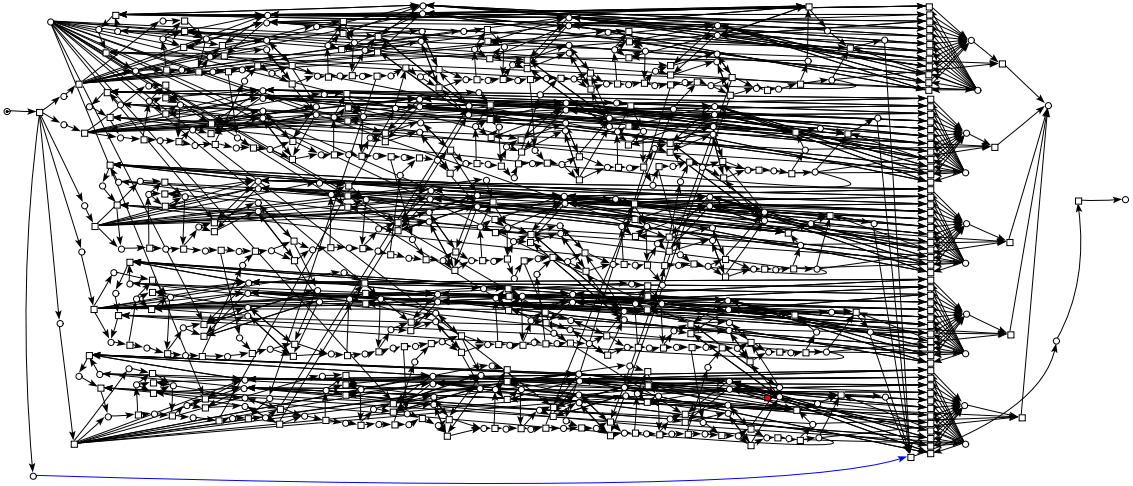} \\ \\
    (b)
\end{tabular}

\end{center}

\caption{Petri nets for Dining Philosophers \textsf{Hash} Solutions}

\label{fig:all_phil_translation}
\end{figure}

Figure \ref{fig:one_phil_translation} presents the Petri nets that
model the respective behaviors of individual philosophers in the
first and in the second solutions. Figure
\ref{fig:all_phil_translation} presents the Petri nets modelling
the interaction among the five philosophers, after modelling
communication channels. The figures are only illustrative, since
the networks have a number of components intractable by simple
visual inspection.

\subsubsection{Proving Properties on the Dining Philosophers Solution}

In this section, INA (Integrated Network Analyzer) is used as an
underlying engine for verifying formal properties about the above
\textsf{Hash} solutions for the dining philosophers problem. INA allows to
perform several structural and dynamic analysis on the Petri net
induced from the two solutions. Among other possible analysis
approaches, INA provides model checking facilities that allows to
check validity of CTL (Constructive Tree Logic) formulae,
describing properties about the \textsf{Hash} program, on the reachability
graph of its corresponding Petri net. CTL is a suitable formalism,
from branching-time temporal logics, for expressing and verifying
safety (invariance) and liveness properties of dynamic systems. It
allows temporal operators to quantify over paths that are possible
from a given state. There exist a superset of CTL, named CTL$^*$,
that augments expressive power of CTL by allowing to express
fairness constrains that are not allowed to be expressed in CTL.
However, INA restricts to CTL because model checking algorithms
for CTL are more efficient (linear in the formula size) that in
CTL$^*$ (exponential in the formula size).

Now, let us to introduce relevant properties that may be proved
about solutions to the dining philosophers problems and to model
these properties using CTL. The dining philosophers problem may be
thought as an instance of the critical section problem. In fact,
forks are critical sections, since it cannot be taken by more than
one philosopher. A good solution to an instance of the critical
section problem must ensure three properties \cite{Andrews1991}:

\begin{itemize}

\item {\bf Mutual exclusion}: Two adjacent philosophers cannot
obtain the same fork (enter critical section);

\item {\bf Absence of deadlock}: A deadlock occurs whenever there
is at least one active philosopher and all active (not terminated)
philosophers are blocked. The classical deadlock situation in
dining philosophers problem occurs when all philosophers acquire
its right, or left, forks. In this state, all philosophers may not
proceed and they are not finished. Thus, they are in deadlock;

\item {\bf Absence of unnecessary delay}: If a philosopher demands
its right (left) fork and their right (left) neighbor is thinking,
the philosopher is not prevented from obtaining the fork;

\item {\bf Eventual entry}: Each philosopher that demands for a
fork eventually will obtain it.

\end{itemize}

\begin{table}
\centering
\begin{center}
\begin{minipage}{14cm}
\begin{footnotesize}
\begin{tabular}{|lcl|c|c|c|}

\hline
\textbf{(a) Macros About Channels} & & &s&b&r \\
\hline
$\textsc{sender\_prepared}[c]$   & $=$ & $\textsc{port\_prepared}[\partial_0(c)]$    &$\bullet$&$\bullet$&$\bullet$\\
$\textsc{receiver\_prepared}[c]$ & $=$ & $\textsc{port\_prepared}[\partial_1(c)]$    &$\bullet$&$\bullet$&$\bullet$ \\
$\textsc{rendezvous}[c]$      & $=$ & $\textsc{sender\_ready}[c] \wedge \textsc{receiver\_ready}[c]$ &$\bullet$&$\bullet$&$\bullet$\\
$\textsc{buffer\_full}[c]$    & $=$ & $\neg\textsc{chan\_buffer\_free}[c]$ &&$\bullet$&\\
$\textsc{buffer\_empty}[c]$   & $=$ & $\neg\textsc{chan\_buffer\_used}[c]$ &&$\bullet$&\\
$\textsc{sender\_blocked}[c]$ & $=$ & $\textsc{sender\_ready}[c] \wedge \neg \textsc{receiver\_ready}[c]$   & $\bullet$&&\\
$\textsc{sender\_blocked}[c]$ & $=$ & $\textsc{sender\_ready}[c] \wedge \textsc{buffer\_full}[c]$           & &$\bullet$&\\
$\textsc{receiver\_blocked}[c]$ & $=$ & $\textsc{receiver\_ready}[c] \wedge \neg \textsc{sender\_ready}[c]$ & $\bullet$&&\\
$\textsc{receiver\_blocked}[c]$ & $=$ & $\textsc{receiver\_ready}[c] \wedge \textsc{buffer\_empty}[c]$ & &$\bullet$& \\

\hline
\end{tabular}
\end{footnotesize}
\end{minipage}
\end{center}

\begin{center}
\begin{minipage}{14cm}
\begin{footnotesize}
\begin{tabular}{|lcl|c|c|}

\hline
\textbf{(b) Macros About Ports} & & &i&o \\
\hline
$\textsc{port\_pair\_prepared}[p]$   & $=$ & $\textsc{port\_prepared}[\partial_1 \circ \partial_0^-1(p)]$   &$\bullet$& \\
$\textsc{port\_pair\_prepared}[p]$   & $=$ & $\textsc{port\_prepared}[\partial_0 \circ \partial_1^-1(p)]$   &         &$\bullet$ \\
\hline

\end{tabular}
\end{footnotesize}
\end{minipage}
\end{center}

\begin{center}
\begin{minipage}{14.0cm}
\begin{footnotesize}
\begin{tabular}{|lcl|c|c|}

\hline
\textbf{(c) Macros About Groups of Ports} & & &all&any \\
\hline
$\textsc{group\_prepared}[G]$ & $=$ & $(\forall p)_{\in G}: \textsc{port\_prepared}[p]$   &$\bullet$&         \\
$\textsc{group\_prepared}[G]$ & $=$ & $(\exists p)_{\in G}: \textsc{port\_prepared}[p]$   &         &$\bullet$\\
\hline

\end{tabular}
\end{footnotesize}
\end{minipage}
\end{center}

\caption{Some Useful Formula Macros for \textsf{Hash} programming}
\label{tbl:ctl_formula_macros_1}
\end{table}

In the following paragraphs, the above properties are
characterized using CTL formulae. But before, intending to
facilitate concise and modular specification of complex CTL
formulas, we define the notion of CTL-formula macro. A CTL-formula
macro is new kind of CTL-formula of the form $\langle macro\_name
\rangle[q_1,q_2,\dots,q_n]$, where $\langle macro\_name \rangle$
is a macro name, $q_i$, $1 \leq i \leq n$, are qualifiers.
CTL-formulae macros may be expanded in flat CTL-formulae, by
applying recursively their definitions. For instance, A
CTL-formula macro is defined using the following syntax:

\begin{center}

$\langle macro\_name \rangle[v_1,v_2,\dots,v_n] :: f$

\end{center}

where $\langle macro\_name \rangle$ is a name for the macro,
$v_i$, $1 \leq i \leq n$, are \emph{qualifier variables}, and $f$
is a CTL-formula (possibly making reference to CTL-formula
macros). In a flat CTL-formula that appears in the right-hand-side
of a CTL-formula macro, qualifier variables are used as qualifiers
for place and transition identifiers and for references to
enclosed CTL-formulae macros.

\begin{table}

\begin{footnotesize}
\begin{tabular}{|lcl|}

\hline
\textbf{(a) Macros About \underline{Philosophers}} & & \textbf{($p=0 \dots N-1$)}\\
\hline
$\textsc{phil\_demands\_lf}[p]$  &=& \\
$\textsc{phil\_demands\_rf}[p]$  &=& \\
$\textsc{phil\_posseses\_lf}[p]$ &=& \\
$\textsc{phil\_posseses\_rf}[p]$ &=& \\
$\textsc{phil\_is\_eating}[p]$   &=& $\textsc{phil\_posseses\_lf}[p] \wedge \textsc{phil\_posseses\_rf}[p]$ \\
$\textsc{phil\_is\_thinking}[p]$ &=& $\neg \textsc{phil\_is\_eating}[p]$\\
$\textsc{phil\_waiting}[p]$      &=& $\textsc{phil\_demands\_rf}[p] \vee \textsc{phil\_demands\_lf}[p]$ \\
$\textsc{phil\_finished}[p]$     &=& $\textsc{process\_finished}[phil[p]] $ \\
$\textsc{all\_phil\_finished}$     &=& $(\forall p)_{< N}: \textsc{phil\_finished}[p] $ \\

\hline

\end{tabular}
\end{footnotesize}

\begin{footnotesize}
\begin{tabular}{|lcl|}
\hline
\textbf{(b) Macros About \underline{Forks}} & & \textbf{($f=0 \dots N-1$)}\\
\hline
$\textsc{fork\_is\_free}[f]$            &=& $\textsc{phil\_is\_thinking}[f] \wedge \textsc{phil\_is\_thinking}[f{\oplus}_{N}1]$ \\
$\textsc{fork\_in\_use\_by\_right}[f]$  &=& $\textsc{phil\_posseses\_lf}[f]$ \\
$\textsc{fork\_in\_use\_by\_left}[f]$   &=& $\textsc{phil\_posseses\_rf}[f{\oplus}_{N}1]$ \\
$\textsc{fork\_in\_use}[f]$             &=& $\textsc{fork\_in\_use\_by\_right}[f] \vee \textsc{fork\_in\_use\_by\_left}[f]$ \\

\hline

\end{tabular}
\end{footnotesize}

\caption{Some Useful Formula Macros for Dining Philosophers}

\label{tbl:ctl_formula_macros_2}

\end{table}

In Table \ref{tbl:ctl_formula_macros_1}, some useful CTL-formulae
macros are defined for simplifying specification of CTL formulae
on the restrict domain of \textsf{Hash} programs. In Table
\ref{tbl:ctl_formula_macros_2}, other CTL-formulae macros are
defined, but now on the restrict domain of the dining philosophers
problem (application oriented). Notice that CTL-formulae macros of
Table \ref{tbl:ctl_formula_macros_2} are defined on top of that
defined in Table \ref{tbl:ctl_formula_macros_1}. Unlike the later
macros, the implementation of the former ones is not sensitive to
modifications in the underlying translation schema. This
illustrates the transparency provided by the use of CTL-formulae
macros in an environment for proof and analysis of formal
properties.

We have used INA for proving the three properties enunciated above
for dining philosophers. The first three ones are \emph{safety
properties}. It may be proved by negating a predicate describing a
state that cannot be reached in execution (\emph{bad state}). The
last one is a \emph{liveness property}, for which validity of a
predicate must be checked in all possible states.

\paragraph{Proof of Mutual Exclusion.} Safety property. One valid
formulation for the corresponding \emph{bad state} is:

\begin{footnotesize}
\begin{center}

$\textsc{bad} = \textbf{EF}\left[(\exists f)_{0 \leq f < N}: \textsc{fork\_in\_use\_by\_right}[f] \wedge \textsc{fork\_in\_use\_by\_left}[f]\right] $ 

\end{center}
\end{footnotesize}

\paragraph{Proof of Absence of deadlock.}  Safety property.
One valid formulation for the corresponding \emph{bad state} is:

\begin{footnotesize}
\begin{center}

$\textsc{bad} = \textbf{EF}\left[\left((\forall p)_{0 \leq p < N}: \textsc{phil\_waiting}[p] \vee \textsc{phil\_finished}[p]\right) \wedge \neg\textsc{all\_phil\_finished}\right] $ 

\end{center}
\end{footnotesize}

\paragraph{Proof of Absence of unnecessary delay.} Safety property. One valid formulation for the corresponding \emph{bad state} is:

\begin{footnotesize}
\begin{center}

$\textsc{bad} = \textbf{EF}\left[\left((\forall p)_{0 \leq f < N}: \textsc{fork\_is\_free}[f] \wedge \left(\textsc{phil\_demands\_lf}[f] \vee \textsc{phil\_demands\_rf}[f{\oplus}_{N}1]\right)\right) \right] $ 

\end{center}
\end{footnotesize}

\paragraph{Proof of Eventual entry.}  Liveness property. One valid formulation for the corresponding \emph{good state} is:
 :

\begin{footnotesize}
\begin{center}

$\textsc{good} = \textbf{AG} \left[\textsc{phil\_waiting}[p] \Rightarrow  \left(\textbf{AF} \left[ \textsc{phil\_eating}[p] \right]\right) \right] $ 

\end{center}
\end{footnotesize}

\subsection{The Alternating Bit Protocol}

The alternating bit protocol (ABP) is a simple and effective
technique for managing retransmission of lost messages in
fault-tolerant low level implementations of message passing
libraries. Given a transmitter process $A$ and a receiver process
$B$, connected by a point-to-point stream channel, ABP ensures
that whenever a message sent from $A$ to $B$ is lost, it is
retransmitted.

\begin{figure}
\begin{center}
    \screendump{0.70}{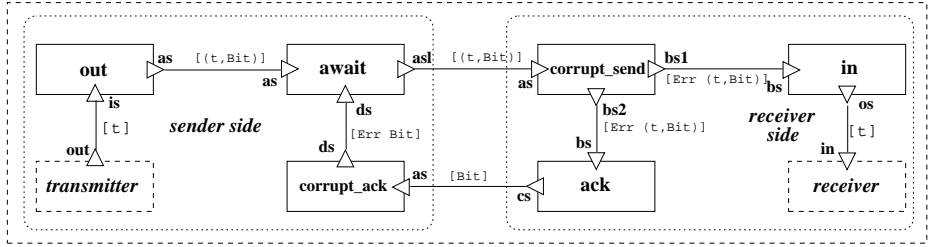}
\caption{\textsf{Hash} Topology for of ABP (Alternating Bit
Protocol)} \label{ABP_diagram}
\end{center}
\end{figure}

The \textsf{Hash} implementation described here is based on a functional
implementation described in \cite{Dybjer1989}. The Figure
\ref{ABP_diagram} illustrates the topology of the component
\textsc{ABP}, which might be used for implementing ABP protocol.
The virtual units \emph{transmitter} and \emph{receiver} model the
processes involved in the communication. The other units implement
the protocol. The units {\em transmitter}, {\em out}, {\em await},
and {\em corrupt\_ack} implement the sender side of the ABP
protocol, while the units {\em receiver}, {\em in}, {\em ack}, and
{\em corrupt\_send} implement the receiver side. The \emph{await}
process may retransmit a message repetitively until the message is
received by process \emph{ack}. Retransmissions are modelled using
streams with nesting factor 2 (streams of streams). The elements
of the nested stream correspond to the retransmission attempts of
a given value. The correct arrive of the message is performed by
inspecting the value received through the port $ds$. The processes
\emph{corrupt\_ack} and \emph{corrupt\_send} verify the occurrence
of errors in the messages that arrive at the sender and receiver,
respectively, modelling the unreliable nature of the communication
channel. The \textsf{Hash} configuration code presented in Figure
\ref{abp_code} implements the \textsc{ABP} component.

\begin{figure}
\begin{center}
\begin{tiny}
\begin{minipage} {\textwidth}
\begin{tabbing}

\textbf{component} \textsc{ABP} \textbf{with} \\
\\
\textbf{use} Out, Await, Corrupt, Ack, In \\
\\
\textbf{interface} \emph{ABP\_Transmitter} \textbf{where} \= \textbf{ports}:  () $\rightarrow$ (out*::t)\\
                                                        \> \textbf{protocol}: \textbf{repeat} out! \textbf{until} out \\
\\
\textbf{interface} \emph{ABP\_Receiver} t \textbf{where} \= \textbf{ports}: (in*::t) $\rightarrow$ () \\
                                                       \> \textbf{protocol}: \textbf{repeat} in? \textbf{until} in  \\
\\
\textbf{interface} \emph{Out} \textbf{where} \= \textbf{ports}: (is::t) $\rightarrow$ (as::(t,Bit))\\
                                 \> \textbf{protocol}: \textbf{repeat} \textbf{seq} \{is?; as!\} \textbf{until} $<$is \& as$>$\\
\\
\textbf{interface} \emph{Await} \textbf{where} \= \textbf{ports}:  (as*::(t,Bit),ds**::Err Bit) $\rightarrow$ (as'**::(t,Bit)) \\
                                               \> \textbf{protocol}: \textbf{repeat} \textbf{seq} \{ as?; \textbf{repeat} \textbf{seq} \{as'!; ds?\} \textbf{until} $<$as' \& ds$>$\} \textbf{until} $<$as \& as' \& ds$>$ \\
\\
\textbf{interface} \emph{Corrupt} \textbf{where} \= \textbf{ports}:   (as**::(t,Bit)) $\rightarrow$ (bs**::Err (t,Bit)) \\
                                                 \> \textbf{protocol}: \textbf{repeat} \textbf{seq} \{as?; bs!\} \textbf{until} $<$as \& bs$>$ \\
\\
\textbf{interface} \emph{Ack} \textbf{where} \= \textbf{ports}: (bs**::Err (t,Bit)) $\rightarrow$ (cs**::Bit) \\
                                             \> \textbf{protocol}: \textbf{repeat} \textbf{seq} \{bs?; cs!\} \textbf{until} $<$bs \& cs$>$ \\
\\
\textbf{interface} \emph{In} \textbf{where} \= \textbf{ports}: (bs**::Err (t,Bit)) $\rightarrow$ (os*::t) \\
                                            \> \textbf{protocol}: \textbf{repeat} \textbf{seq} \{ \textbf{repeat} bs? \textbf{until} bs; os!\} \textbf{until} $<$bs \& os$>$\\
\\
\textbf{unit} transmitter  \= \textbf{where ports}: \emph{ABP\_Transmitter} () $\rightarrow$ out\\
\textbf{unit} receiver     \> \textbf{where ports}: \emph{ABP\_Receiver} in $\rightarrow$ ()\\
\\
\textbf{unit} out           \ \ \ \ \ \ \ \ \ \ \ \ \= \textbf{where ports}: \emph{IOut}     \hspace*{2.8cm}\= \textbf{assign} {Out} \hspace*{0.3cm}     \= \textbf{to} out\\
\textbf{unit} await                                 \> \textbf{where ports}: \emph{IAwait}                  \> \textbf{assign} {Await}   \> \textbf{to} await \\
\textbf{unit} corrupt\_ack                          \> \textbf{where ports}: \emph{ICorrupt}                \> \textbf{assign} {Corrupt} \> \textbf{to} corrupt\_ack\\
\textbf{unit} in                                    \> \textbf{where ports}: \emph{IIn}                     \> \textbf{assign} {In}      \> \textbf{to} in \\
\textbf{unit} ack                                   \> \textbf{where ports}: \emph{IAck}                    \> \textbf{assign} {Ack}     \> \textbf{to} ack \\
\textbf{unit} corrupt\_send                         \> \textbf{where ports}: \emph{ICorrupt} \textbf{grouping}: bs*2 \textbf{all} \> \textbf{assign} {Corrupt} \> \textbf{to} corrupt\_send \\
\\
\textbf{connect *} transmitter$\rightarrow$out     \ \ \ \ \ \ \= \textbf{to} out$\leftarrow$is           \ \ \ \ \ \ \ \ \ \ \ \ \ \ \=   \\
\textbf{connect *} in$\rightarrow$os               \ \ \ \ \ \ \> \textbf{to} receiver$\leftarrow$in      \ \ \ \ \ \ \ \ \ \ \ \ \ \ \>   \\
\textbf{connect *} out$\rightarrow$as              \ \ \ \ \ \ \> \textbf{to} await$\leftarrow$as         \ \ \ \ \ \ \ \ \ \ \ \ \ \ \>  \\
\textbf{connect *} await$\rightarrow$as            \ \ \ \ \ \ \> \textbf{to} corrupt\_send$\leftarrow$as \ \ \ \ \ \ \ \ \ \ \ \ \ \ \>, \textbf{buffered}  \\
\textbf{connect *} corrupt\_ack$\rightarrow$ds     \ \ \ \ \ \ \> \textbf{to} ack$\leftarrow$ds           \ \ \ \ \ \ \ \ \ \ \ \ \ \ \>  \\
\textbf{connect *} corrupt\_send$\rightarrow$bs[0] \ \ \ \ \ \ \> \textbf{to} in$\leftarrow$bs            \ \ \ \ \ \ \ \ \ \ \ \ \ \ \>  \\
\textbf{connect *} corrupt\_send$\rightarrow$bs[1] \ \ \ \ \ \ \> \textbf{to} ack$\leftarrow$bs           \ \ \ \ \ \ \ \ \ \ \ \ \ \ \>  \\
\textbf{connect *} ack$\rightarrow$cs              \ \ \ \ \ \ \> \textbf{to} corrupt\_ack$\leftarrow$cs  \ \ \ \ \ \ \ \ \ \ \ \ \ \ \>, \textbf{buffered}  

\end{tabbing}
\end{minipage}
\end{tiny}
\caption{\textsc{ABP} Component} \label{abp_code}
\end{center}
\end{figure}

\begin{figure}
\begin{center}
\begin{minipage}{\textwidth}
    \centering
    \screendump{0.4}{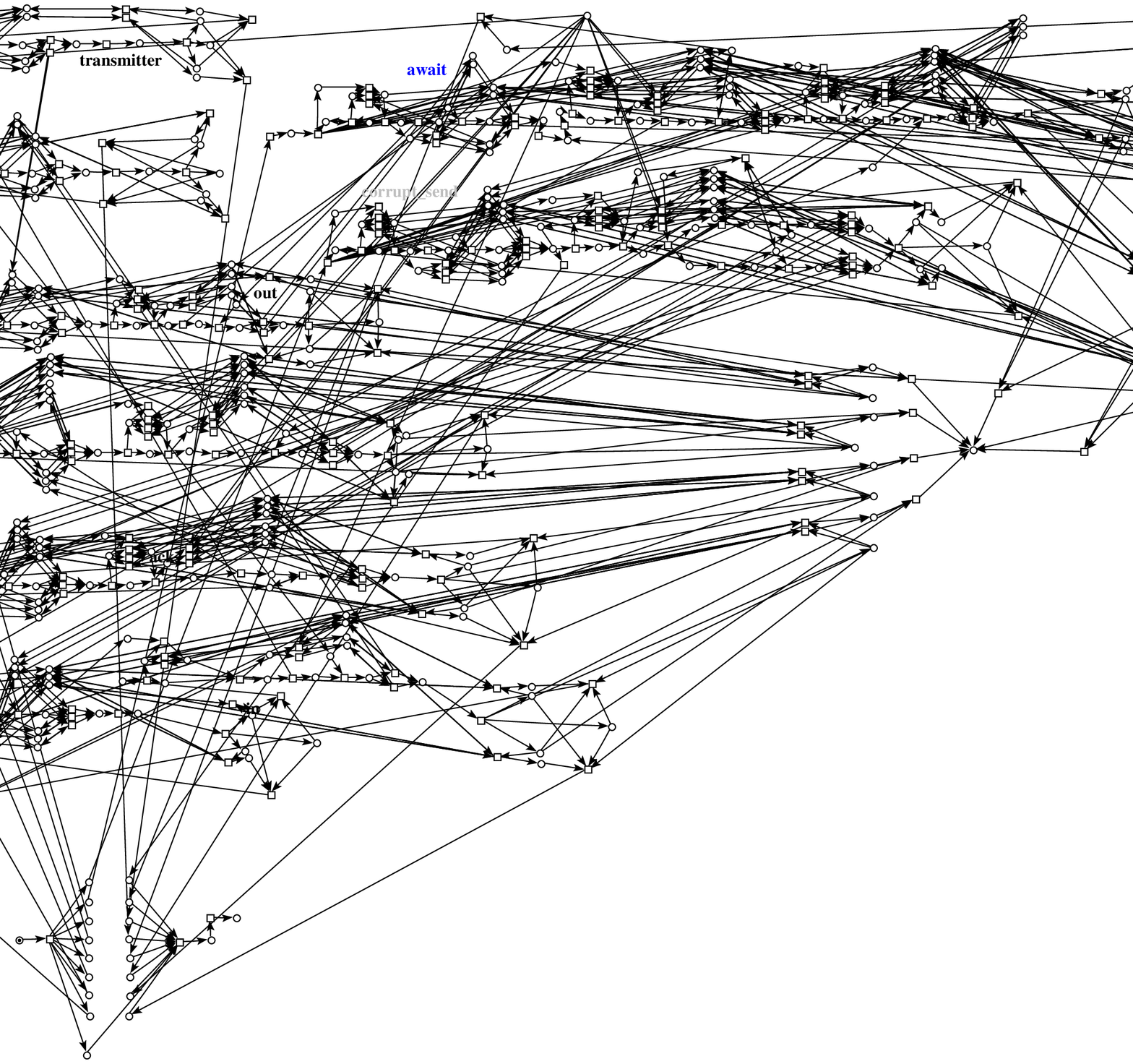}
\end{minipage}

\end{center}

\caption{Petri Net Induced by \textsc{ABP} Component}

\label{abp_translation_1}
\end{figure}

The Petri net induced by translating \textsc{ABP} component is
presented in Figure \ref{abp_translation_1}.




\appendix

\section{The Formal Syntax of HCL}
\label{ap:hcl_syntax}

In what follows, it is described a context-free grammar for HCL,
the Haskell$_\#$ Configuration Language, whose syntax and
programming abstractions were informally presented in Section
\ref{sec2}. Examples of HCL configurations and their meanings were
presented in Sections \ref{sec2} and \ref{sec3}. The notation
employed here is similar to that used for describing syntax of
Haskell 98 \cite{PeytonJones99}. Indeed, some non-terminals from
that grammar are reused here, once some Haskell code appears in
HCL configurations. They are faced italic and bold. A minor
difference on notation resides on the use of $(\dots)^?$, instead
of $[\dots]$, for describing optional terms. For simplicity,
notation for indexed notation is ignored from the description of
formal syntax of HCL. It may be resolver by a pre-processor,
before parsing.

\subsection{Top-Level Definitions}

\begin{normalsize}
\begin{tabbing}
\emph{configuration} \hspace{1.0cm} \= $\rightarrow$ \emph{header} \emph{declaration}$_1$ $\dots$ \emph{declaration}$_n$ ($n \geq 0$)\\
\emph{header} \> $\rightarrow$ \texttt{component} ID \emph{static\_parameter\_list}$^?$ \emph{component\_interface}$^?$ \\
\emph{static\_parameter\_list} \> $\rightarrow$ $<$ ID$_1$ $\dots$ ID$_n$ $>$  ($n \geq 0$)\\
\emph{component\_interface} \> $\rightarrow$ \emph{ports\_naming} \\
\emph{declaration} \> $\rightarrow$ \= \emph{import\_decl} \ \= $\mid$ \emph{use\_decl} \ \ \ \ \ \ \ \= $\mid$ \emph{iterator\_decl} \ \= $\mid$ \emph{interface\_decl} \\
                   \>\ $\mid$       \> \emph{unit\_decl}     \> $\mid$ \emph{assign\_decl}      \> $\mid$ \emph{replace\_decl}    \> $\mid$ \emph{channel\_decl}  \\
                   \>\ $\mid$       \> \emph{unify\_decl}    \> $\mid$ \emph{factorize\_decl}   \> $\mid$ \emph{replicate\_decl}       \> $\mid$ \emph{bind\_decl} \\
                   \>\ $\mid$       \> \emph{haskell\_code} 
\end{tabbing}
\end{normalsize}

\subsection{Use Declaration}

\begin{normalsize}
\begin{tabbing}
\emph{use\_decl} $\rightarrow$ \texttt{use} \emph{use\_spec}  \\
\emph{use\_spec} $\rightarrow$ id $\mid$ id.\emph{use\_spec} $\mid$ id.\{ \emph{use\_spec}$_1$ , $\dots$ , \emph{use\_spec}$_n$ \}\hspace{1.0cm} ($n \geq 1$) 

\end{tabbing}
\end{normalsize}

\subsection{Import Declaration}

\begin{normalsize}
\begin{tabbing}
\emph{import\_decl} $\rightarrow$ \textbf{\emph{impdecl}}

\end{tabbing}
\end{normalsize}

\subsection{Iterator Declaration}

\begin{normalsize}
\begin{tabbing}

\emph{iterator\_decl} $\rightarrow$ \texttt{iterator} id$_1$, $\dots$, id$_n$ \texttt{range} [ \emph{numeric\_exp} , \emph{numeric\_exp} ]  ($n \geq 1$)

\end{tabbing}
\end{normalsize}

\subsection{Interface Declaration}

\begin{normalsize}
\begin{tabbing}
\emph{interface\_decl} \= $\rightarrow$ \texttt{interface} (\emph{\textbf{context}} $=>$)$^?$ ID \textbf{\emph{tyvar}}$_1$ $\dots$ \textbf{\emph{tyvar}}$_k$ \emph{interface\_spec} \\
\emph{interface\_spec} \> $\rightarrow$ \= \emph{interface\_ports\_spec} \\
                       \>               \> (\texttt{where} : \emph{interface\_inheritance})$^?$ (\texttt{behavior} : \emph{behavior\_expression})$^?$ 

\end{tabbing}
\end{normalsize}

\subsubsection{Interface Ports Description}

\begin{normalsize}
\begin{tabbing}
\emph{interface\_ports\_spec} \= $\rightarrow$ \emph{port\_spec\_list} -$>$ \emph{port\_spec\_list} \\
\emph{port\_spec\_list} \> $\rightarrow$ \emph{port\_spec} $\mid$ ( \emph{port\_spec}$_1$ , $\dots$ , \emph{port\_spec}$_n$ ) ($n \geq 2$)\\
\emph{port\_spec} \> $\rightarrow$ id (*)$^?$ (:: \emph{\textbf{atype}})$^?$ $\mid$ id 

\end{tabbing}
\end{normalsize}

\subsubsection{Interface Composition}

\begin{normalsize}
\begin{tabbing}
\emph{interface\_inheritance} \hspace*{0.7cm} \= $\rightarrow$ \emph{interface\_slice}$_1$ \# $\dots$ \# \emph{interface\_slice}$_k$  ($k \geq 1$) \\
\emph{interface\_slice} \> $\rightarrow$ id @ ID $\mid$ ID \emph{ports\_naming\_composition} \\
\emph{ports\_naming\_composition} \> $\rightarrow$ \= \emph{ports\_naming} \\
                                             \>\ $\mid$        \> ( \emph{ports\_naming}$_1$ \# $\dots$ \# \emph{ports\_naming}$_n$)\ \ \ ($n \geq 1$)\\
\emph{ports\_naming} \> $\rightarrow$ \emph{port\_naming\_list} -$>$ \emph{port\_naming\_list} \\
\emph{port\_naming\_list} \> $\rightarrow$ id $\mid$ ( id$_1$ , $\dots$ , id$_n$)   ($n \geq 1$) 
\end{tabbing}
\end{normalsize}

\subsubsection{Interface Behavior}

\begin{normalsize}
\begin{tabbing}
\\
\emph{behavior\_expression}\ \ \= $\rightarrow$ (\texttt{sem} id$_1$ , $\dots$ , id$_n$)$^?$ : \emph{action} \ \ \ ($n \geq 1$) \\
\emph{action}  \> $\rightarrow$ \= \texttt{par}  \{ \emph{action}$_1$ ; $\dots$ ; \emph{action}$_n$ \} $\mid$ \texttt{seq} \{ \emph{action}$_1$ ; $\dots$ ; \emph{action}$_n$ \} \\
               \>\ $\mid$       \> \texttt{alt}  \{ \emph{action}$_1$ ; $\dots$ ; \emph{action}$_n$ \} $\mid$ \texttt{repeat} \emph{action} \emph{condition}$^?$ \\
               \>\ $\mid$       \> \texttt{if} \emph{condition} \texttt{then} \emph{action} \texttt{else} \emph{action}  \\
               \>\ $\mid$       \> id ! $\mid$ id ? $\mid$ \texttt{signal} id $\mid$ \texttt{wait} id\ \ \ \ ($n \geq 2$) \\
\emph{condition} \> $\rightarrow$ \texttt{until} \emph{disjunction} $\mid$ \texttt{counter} \emph{numeric\_exp} \\
\emph{disjunction} \> $\rightarrow$ \emph{sync\_conjunction}$_1$ `$\mid$' $\dots$ `$\mid$' \emph{sync\_conjunction}$_n$\ \ \  ($n \geq 1$)\\
\emph{sync\_conjunction} \> $\rightarrow$ $\langle$ \emph{simple\_conjunction} $\rangle$ $\mid$ \emph{simple\_conjunction} \\
\emph{simple\_conjunction} \> $\rightarrow$ id $\mid$ ( id$_1$ \& $\dots$ \& id$_n$ )\ \ \   ($n \geq 1$) 

\end{tabbing}
\end{normalsize}

\subsection{Unit Declaration}

\begin{normalsize}
\begin{tabbing}
\emph{unit\_decl} \hspace*{0.6cm} \= $\rightarrow$ \texttt{unit} \emph{unit\_spec} \\
\emph{unit\_spec} \> $\rightarrow$ (*)$^?$ id (\# \emph{unit\_interface})$^?$ (\texttt{wire} \emph{wf\_setup}$_1$ , $\dots$ , \emph{wf\_setup}$_n$)$^?$ \\
\emph{unit\_interface} \> $\rightarrow$ ID \emph{ports\_naming\_composition}$^?$ $\mid$ \emph{interface\_spec} \\
\emph{wf\_setup} \> $\rightarrow$ \= id (\emph{group\_type} \emph{group\_spec})$^?$ (: \emph{wire\_function})$^?$ \\ 
\emph{group\_spec} \> $\rightarrow$ \{ id$_1$, $\dots$, id$_n$ \}  $\mid$ * \emph{numeric\_exp}  \\
\emph{group\_type} \> $\rightarrow$ \texttt{any} $\mid$ \texttt{all} \\
\emph{wire\_function} \> $\rightarrow$ ? $\mid$ \emph{\textbf{exp}} 

\end{tabbing}
\end{normalsize}

\subsection{Assignment Declaration}

\begin{normalsize}
\begin{tabbing}
\emph{assign\_decl} \hspace*{1.4cm} \= $\rightarrow$ \texttt{assign} \emph{assigned\_component} \texttt{to} \emph{assigned\_unit} \\
\emph{assigned\_component} \> $\rightarrow$ ID \emph{actual\_parameter\_list}$^?$ \emph{ports\_naming\_composition}$^?$ \\
\emph{actual\_parameter\_list} \> $\rightarrow$ $<$ \emph{numeric\_exp}$_1$ , $\dots$ , \emph{numeric\_exp}$_n$ $>$ ($n \geq 1$) \\
\emph{assigned\_unit} \> $\rightarrow$ \emph{qid} \emph{ports\_naming\_composition}$^?$ 

\end{tabbing}
\end{normalsize}

\subsection{Replace Declaration}

\begin{normalsize}
\begin{tabbing}
\emph{replace\_decl} \= $\rightarrow$ \texttt{replace} \emph{qid} \emph{ports\_naming\_composition}$^?$ \texttt{by} \emph{operand\_unit} 

\end{tabbing}
\end{normalsize}

\subsection{Channel Declaration}

\begin{normalsize}
\begin{tabbing}
\emph{channel\_decl} \= $\rightarrow$ \texttt{connect} \emph{qid} -$>$ \emph{qid} \texttt{to} \emph{qid} $<$- \emph{qid} , \emph{comm\_mode} \\
\emph{comm\_mode} \> $\rightarrow$ \texttt{synchronous} $\mid$ \texttt{buffered} numeric\_exp $\mid$ \texttt{ready} 

\end{tabbing}
\end{normalsize}

\subsection{Unification Declaration}

\begin{normalsize}
\begin{tabbing}
\emph{unify\_decl} \hspace{1.0cm} \= $\rightarrow$ \texttt{unify} \= \emph{operand\_unit}$_1$ , $\dots$ , \emph{operand\_unit}$_n$ \texttt{to} \emph{unit\_spec} \\
                                  \>                              \> \texttt{adjust} \texttt{wire} \emph{wf\_setup}$_1$ , $\dots$ , \emph{wf\_setup}$_k$\ \ \ ($n \geq 2,\ k \geq 1$)\\
\emph{operand\_unit} \> $\rightarrow$ \emph{qid} \# \emph{interface\_pattern}$_1$ $\dots$ \# \emph{interface\_pattern}$_n$ ($n \geq 1$) \\
\emph{interface\_pattern} \> $\rightarrow$ \emph{port\_pattern\_list} -$>$ \emph{port\_pattern\_list} $\mid$ id \\
\emph{port\_pattern\_list} \> $\rightarrow$  \emph{pattern} $\mid$ ( \emph{pattern}$_1$ , $\dots$ , \emph{pattern}$_n$ ) \\
\emph{pattern} \> $\rightarrow$ id $\mid$ @ \emph{qid} $\mid$ \_ $\mid$ \_\_ 
\end{tabbing}
\end{normalsize}

\subsection{Factorization Declaration}

\begin{normalsize}
\begin{tabbing}
\emph{factorize\_decl} $\rightarrow$ \texttt{factorize} \= \emph{operand\_unit} \texttt{to} \emph{unit\_spec}$_1$ $\dots$ \emph{unit\_spec}$_n$  \\
                                                          \> \texttt{adjust} \texttt{wire} \emph{wf\_setup}$_1$ , $\dots$ , \emph{wf\_setup}$_k$\ \ \ ($n \geq 2,\ k \geq 1$) 

\end{tabbing}
\end{normalsize}

\subsection{Replication Declaration}

\begin{normalsize}
\begin{tabbing}
\emph{replicate\_decl} $\rightarrow$ \texttt{replicate} \= \emph{operand\_unit}$_1$ , $\dots$ , \emph{operand\_unit}$_n$ \texttt{into} \emph{numeric\_exp} \\
                                                          \> \texttt{adjust} \texttt{wire} \emph{wf\_setup}$_1$ , $\dots$ , \emph{wf\_setup}$_k$\ \ \ ($n \geq 2,\ k \geq 1$) 

\end{tabbing}
\end{normalsize}

\subsection{Bind Declaration}

\begin{normalsize}
\begin{tabbing}
\emph{bind\_declaration} $\rightarrow$ \texttt{bind} \emph{qid} -$>$ \emph{qid} \texttt{to} -$>$ id $\mid$ \texttt{bind} \emph{qid} $<$- \emph{qid} \texttt{to} $<$- id 

\end{tabbing}
\end{normalsize}

\subsection{Miscelaneous}

\begin{normalsize}
\begin{tabbing}
\emph{haskell\_code} $\rightarrow$ \emph{\textbf{topdecls}} \\
\\
\emph{qid} $\rightarrow$ id$_1$ `.' $\dots$ `.' id$_n$\ \ \ ($n \leq 2$)\\
\emph{qID} $\rightarrow$ ID$_1$ `.' $\dots$ `.' ID$_n$\ \ \ ($n \leq 2$)\\

\end{tabbing}
\end{normalsize}

\section{Foundations and Notations}
\label{sec:preliminaries_and_notation}

In this section, it is discussed the formalisms that comprise the
formal framework for the development of this work, concerning
modelling of communication behavior of processes, according to the
\textsf{Hash} component model design principles.

\subsection{Formal Languages}

The theory of formal languages will be employed as a framework for
the study of patterns of communication interaction of \textsf{Hash} process
in a parallel program. The main interest is to investigate
relations between descriptive power of concurrent expressions and
Petri nets, in order to define a language for expressing
communication behavior of processes, embedded in the \textsf{Hash} language.

\begin{definition} [Alphabet]

An alphabet is a finite set of indivisible symbols, denoted by
$\Sigma$.

\end{definition}

\begin{definition} [Word]

A word is a finite sequence of symbols of some alphabet $\Sigma$.
The symbol $\epsilon$ denotes the empty word, whose length is
zero.

\end{definition}

\begin{definition} [Kleene's Closure of an Alphabet ]

A Kleene's closure of an alphabet $\Sigma$, denoted  by
$\Sigma^*$, is defined as below:

\begin{center}

$\Sigma^* = \{ w \mid w\ is\ a\ word\ in\ \Sigma \}$

\end{center}

Thus, any sequence of symbols in $\Sigma$, including $\epsilon$,
belongs to $\Sigma^*$. It is common to define $\Sigma^+$ as:

\begin{center}

$\Sigma^+ = \Sigma^* - \{ \epsilon \}$

\end{center}

\end{definition}

\begin{definition}
\label{formallanguage_def}

Given an alphabet $\Sigma$, a formal language L is defined as
follows:

\begin{center}

$L \subset \Sigma^*$

\end{center}

\end{definition}

\subsection{Labelled Petri Nets and Formal Languages}

Now, notations and definitions concerning Petri nets are
presented.

\begin{definition}[Place/Transition Petri Net]

A place/transition Petri net is a directed bipartite graph that
can be formalized as a quadruple $(P,T,A,M_0)$, where:

\begin{enumerate}

\item $P$ is a finite set of \emph{places}, which can store an
unlimited number of \emph{marks};

\item $T$ is a finite set of \emph{transitions}.

\item $P \cap T = \emptyset$.

\item $A$ defines a set of \emph{arcs}, in such way that $A
\subseteq ((P \times T) \cup (T \times P)) \times Naturais$. Thus,
an arc can go from a transition to a place or from a place to a
transition. A number is associated to the arc, indicating its
\emph{weight}. For simplicity, if the weight is omitted it is one
($(p,t) \equiv (p,t,1)$).

\item The relation $M_0 \subset P \times N$ defines the
\emph{initial marking}, or the number of marks that are stored in
each place at the initial state of the Petri net;

\end{enumerate}

\label{petrinet_def}
\end{definition}

From the initial marking, a Petri net defines a set reachable
markings. A marking is reachable if it can be obtained from the
initial marking by firing a sequence of transitions, according to
firing rules, formalized in what follows.

\begin{definition} [Enabled Transition]

Be $\Pi$, $\Pi = (P, T, A, M_O)$, a Petri net and $t$, $t \in T$,
a transition of $\Pi$:

\begin{center}

$t \mbox{ is enabled} \Leftrightarrow (\forall p \in P, \exists m,
n \in N: (p, t, m) \in A \wedge (p, n) \in M : n \geq m $)

\end{center}

\end{definition}

Thus, a transition $t$ is enabled if the number of marks in each
one of its input places is greater than or equal to the weight of
the arc that links it to the transition.

\begin{definition}[Firing Rule and Reachable Markings]

Be $\Pi = (P, T, A, M_0)$ a Petri net and a marking $M$ of $\Pi$.
The transition from the marking M to a new marking M' is a side
effect of \emph{firing} an enabled transition $t$, and represented
by the relation $M \stackrel{t}{\rightarrow} M'$, where:

\begin{center}
$M' = \bar{M} \cup M_I \cup M_O$ \\
$M_I = \{ (p,n-m) \mid \exists m, n \in N: p \in P, (p,t,m) \in A, (p,n) \in M\} $ \\
$M_O = \{ (p,n+m) \mid \exists m, n \in N: p \in P, (t,p,m) \in A, (p,n) \in M\} $ 

\end{center}

The sub-marking $M_I$ indicates the new marking for the input
places of the transition $t$, while $M_O$ indicates the new
marking for output places of $t$. The transition can be fire if it
is enabled and the effect of firing is to remove marks from input
places and add marks to output places, according to the weights of
the arcs that link these places to $t$. It is necessary to
generalize this definition to cover the concept of (transitively)
reachable marking. Thus, a marking $M_n$ is reachable from a mark
M if there is a sequence of firing of transitions from M to $M_n$:

\begin{center}
$M \stackrel{t_1}{\rightarrow} M_1 \stackrel{t_2}{\rightarrow} M_2
\stackrel{t_3}{\rightarrow} \cdots \stackrel{t_n}{\rightarrow}
M_n$
\end{center}

This notation can be abbreviated to:

\begin{center}
$M \stackrel{\sigma}{\rightarrow} M_n$, where $\sigma = t_1 t_2
\cdots t_n$
\end{center}

The symbol $\sigma$ denotes a sequence of firings of transitions.

\end{definition}

\begin{definition}[Labelled Petri nets]

Be $\Sigma$ an alphabet. A labelled Petri net is a 7-uple
$(P,T,A,M_0,\rho)$, where $(P,T,A,M_0)$ is a place/transition
Petri net and $\rho:T\rightarrow\Sigma\cup\{\lambda\}$ is a
function that associate transitions to a symbol in $\Sigma$.
Transitions labelled with $\lambda$ are called \emph{silent
transitions}.

\label{redepetrirotulada_def}
\end{definition}

Labelled Petri nets are an extension of Petri nets for generating
formal languages. There are two classes of Petri net languages.

\begin{definition}[Petri Net Language]
Given a labelled Petri net $\Pi = (P,T,A,M_0,\rho)$, we define the
formal language generated by $\Pi$ as:

\begin{center}
$L(\Pi,M_0) = \{\rho(\sigma) \mid \exists M: M_0
\stackrel{\sigma}{\rightarrow} M$\}.
\end{center}

\end{definition}

\begin{definition}[Petri Net Terminal Language]
\label{def:terminal_petri_net_language}

Given a Petri net $\Pi (P,T,A,M_0,\rho)$ and a \emph{final
marking} $M_f$, we can define the terminal language generated by
$\Pi$, with respect to $M_f$, by:

\begin{center}
$T(N,M_0,M_f) = \{\rho(\sigma) \mid M_0
\stackrel{\sigma}{\rightarrow} M_f\}$
\end{center}

\end{definition}

Given a labelled Petri net $\Pi$, the language generated by $\Pi$
defines the possible sequence of firing traces from its initial
marking. The terminal language generated by $\Pi$, with respect to
final marking $M_f$, differs from $\Pi$ language because only that
traces from $M_0$ to $M_f$ are considered.

\begin{definition} [Classes of Petri Nets Languages]

We denote the class of all Petri net languages as
$\ell{_\lambda}{^0}$, and the class of all Petri net terminal
languages as $\ell{_\lambda}{^1}$. It is simple to demonstrable
that $\ell{_\lambda}{^0} \subset \ell{_\lambda}{^1}$.

\end{definition}

\subsection{Interlaced Petri Nets}
\label{sec:interlaced_petri_nets}

Interlaced Petri nets are an alternative extension to labelled
Petri nets, introduced in this article for simplifying
specification of the translation schema of \textsf{Hash} programs into
labelled Petri nets. Like hierarchic Petri nets \cite{},
interlaced Petri nets allows that complex and large scale Petri
nets be implemented in a simpler and modular way. It does not
modify descriptive power of labelled Petri nets. However, while
hierarchic Petri nets allow only nesting composition of Petri
nets, interlaced Petri nets also allow overlapping of them. Given
a set of \emph{Petri net slices}, each one addressing different
\emph{concerns}, they may be overlapped to form an
\emph{interlaced Petri net}. Interlaced Petri nets may also be
viewed as Petri net slices for composing higher level interlaced
Petri nets.

\begin{definition} [Interlaced Petri Nets]

An interlaced Petri net $\Pi$ can be defined inductively as:

\begin{enumerate}

\item (\textbf{Base}) A tuple $(P,T,A,M_0,\rho,\delta)$, where
$(P,T,A,M_0,\rho)$ represents a labelled Petri net and $\delta: (P
\cup T) \rightarrow 2^\Delta $ is a function that maps Petri net
nodes (places and transitions) onto a list of \emph{qualifiers},
represents a \emph{simple} interlaced Petri net;

\item (\textbf{Induction}) Let $\Pi_1,\Pi_2,\dots,\Pi_n$ be
interlaced Petri nets, called \emph{Petri net slices} in this
context (\emph{hypothesis}). $\Pi$, such that $\Pi =
\langle\Pi_1,\Pi_2,\dots,\Pi_n\rangle$, is a \emph{composed}
interlaced Petri net (\emph{induction step}).

\item Anything that may not be formed from application of
inductive rules 1 and 2 is not an interlaced Petri net.

\end{enumerate}

\end{definition}

\begin{definition} [Unfolding Composite Interlaced Petri Nets]

An unfolded interlaced Petri net $\breve{\Pi}$ is obtained from a
interlaced petri net $\Pi$ by applying the transformation function
$\mu$, defined below:

\begin{enumerate}

\item $\mu\left[(P,T,A,M_0,\rho,\delta)\right] =
(P,T,A,M_0,\rho,\delta)$

\item $\mu\left[<\Pi_1,\Pi_2,\dots,\Pi_n>\right] =
\mu\left[\Pi_1\right] \breve{\cup} \mu\left[\Pi_2\right]
\breve{\cup} \dots \breve{\cup} \mu\left[\Pi_n\right]$ (composite)

\end{enumerate}

\end{definition}

The binary operator $\breve{\cup}$ correspond to the Petri net
union operator, defined as below:
\begin{center}
\begin{tiny}
$(P_1,T_1,A_1,{M_0}_1,\rho_1,\delta_1) \breve{\cup}
(P_2,T_2,A_2,{M_0}_2,\rho_2,\delta_2) = (P_1 \cup P_2, T_1 \cup
T_2,A_1 \cup A_2, {M_0}_1 \cup {M_0}_2,\rho_1 \cup \rho_2,\delta_1
\cup \delta_2) $
\end{tiny}
\end{center}

Qualifiers are used to identify components of interlaced Petri
nets (places or transitions) that must be treated as the same
component. These components are said to be \emph{equivalent
components}. They may belong to distinct \emph{slices}. Next,
qualifiers and identification rules are formalized.

\begin{definition} [Qualifier]

Let $\Theta$ be a finite set of symbols. The set $\Delta$ is
informally defined as all qualifiers that can be induced from set
$\Theta$. A qualifier $\delta$, $\delta \in \Delta$, is defined
from $\Theta$ as following:

\begin{itemize}

\item $a$, $a \in \Theta$, is a qualifier (\emph{primitive
qualifier});

\item A tuple $(a_1, a_2, \dots, a_n)$ is a qualifier
(\emph{composed qualifier}), assuming that each $a_i$, $1 \leq i
\leq n$, is a primitive qualifier;


\end{itemize}

\end{definition}

The following rule teaches how to identify equivalent vertices in
an interlaced Petri net, using their qualifiers.

\begin{definition} [Identification of Vertices (Places or Transitions)]

Let $\Pi$ be an interlaced Petri net. Consider its unfolded
variant $\mu\left[\Pi\right] = (P,T,A,M_0,\rho,\delta)$. Let $V$
be the collection of components of $\mu\left[\Pi\right]$ ($P \cup
T$). The following equivalence relation is defined between two
vertices $v_1$ and $v_2$ ($\{v_1, v_2\} \subset V$):

\begin{center}
$v_1 \equiv v_2 \Leftrightarrow (\{v_1,v_2\} \subset T \vee
\{v_1,v_2\} \subset P) \wedge (\delta(v_1) \cap \delta(v_2) \neq
\emptyset) $
\end{center}

\end{definition}

\subsection{Regular Expressions Controlled by Balanced Semaphores
(SCRE)}

The following definitions are presented in order to introduce the
class of regular expressions controlled by balanced semaphores
(SCRE). This generalization to regular expressions, with
descriptive power comparable to labelled Petri nets, will be used
to model a language for specification of communication behavior of
\textsf{Hash} processes.

\begin{definition} [Regular Expressions]

A regular expression E over an alphabet $\Sigma$ is inductively
defined as follows:

\begin{enumerate}

\item a $\in$ $\Sigma$ is a regular expression;

\item $\lambda$ is a regular expression;

\item $\emptyset$ is a regular expression;

\item Be $S_1$ and $S_2$ regular expression. Thus, ($S_1$),
${S_1}\cdot{S_2}$, ${S_1}+{S_2}$ e, ${S_1}^*$ are also regular
expressions.

\end{enumerate}

\end{definition}

\begin{definition}[Regular Language]
\label{regularlanguages_def}

Be $E$ a regular expression over $\Sigma$. The formal language
generated by $E$, $L_{\textsf{RE}}(E)$, is defined in the
following way:

\begin{enumerate}

\item $L_{\textsf{RE}}(\emptyset) = \emptyset$

\item $L_{\textsf{RE}}(\lambda) = \{\lambda\}$

\item $L_{\textsf{RE}}(a) = \{a\},\ para\ a \in \Sigma$,

\item $L_{\textsf{RE}}((E)) = L_{\textsf{RE}}(E)$

\item $L_{\textsf{RE}}({E_1}.{E_2}) = \{xy \mid x \in
L_{\textsf{RE}}(E_1) \wedge y \in L_{\textsf{RE}}(E_2)\}$

\item $L_{\textsf{RE}}(E_1+E_2) = \{x \mid x \in
L_{\textsf{RE}}(E_1) \vee x \in L_{\textsf{RE}}(S_2)\}$

\item $L_{\textsf{RE}}(E^*) = \bigcup L_{\textsf{RE}}(E_i)$

\end{enumerate}

where:

\begin{enumerate}
\item $E^0 = \lambda$ \item $E^i = E^{i-1}E, i > 0$
\end{enumerate}

\end{definition}

The class of regular languages are denoted by $\textsf{RE}$.

A previous version of the \textsf{Hash} language used regular expressions to
model communication traces of processes \cite{Carvalho2002a}.
However, Petri nets are far more expressive than simple regular
expressions for describing communication traces. This fact
motivated us to generalize the adopted approach, using some class
of synchronized concurrent expressions shown to be equivalent to
Petri nets. Concurrent expressions are an extension to regular
expressions defined to model concurrency. Some authors refer to
these languages as \emph{shuffle languages}.

\begin{definition}[Concurrent Expressions]
\label{concexpressions_def}

Be $\Sigma$ an alphabet. A concurrent expression S over $\Sigma$
is defined by the following rules:

\begin{enumerate}

\item If $S$ is a regular expression over $\Sigma$, then $S$ is a
concurrent expression;

\item If $S_1$ e $S_2$ are concurrent expressions, then
${S_1}\odot{S_2}$, e ${S_1}^{\otimes}$ are concurrent expressions;

\end{enumerate}
\end{definition}

The operators $\odot$ and $\otimes$ distinguish, at least
syntactically, non-regular concurrent expressions. The language
generated by concurrent expressions are defined below.

\begin{definition}[Concurrent Languages]
\label{conclanguages_def}

Be $S$ a concurrent expression over $\Sigma$. The concurrent
language $L_{CE}(S)$ is defined according to the following rules:

\begin{enumerate}

\item $L_{\textsf{CE}}(S) = L_{\textsf{RE}}(S)$, if $S$ is a
well-formed regular expression.

\item $L_{\textsf{CE}}(S_1 \odot S_2) =
\{{x_1}{y_1}{x_2}{y_2}\cdots{x_k}{y_k} \mid {x_1}{x_2}\cdots{x_k}
\in L_{\textsf{CE}}(S_1) \wedge {y_1}{y_2}\cdots{y_k} \in
L_{\textsf{CE}}(S_2) \}$,

\item $L_{\textsf{CE}}(S^{\otimes}) = \bigcup
L_{\textsf{CE}}(S_{{\odot}i})$.

\end{enumerate}

where:

\begin{enumerate}
\item $S^{\odot 0} = \lambda$ \item $S^{\odot i} = S^{\odot i-1}
\odot S, i > 0$.
\end{enumerate}

\end{definition}

The class of concurrent languages are denoted by $\textsf{CE}$. It
is clear that $\textsf{CE} \supset \textsf{RE}$, because every
regular expression is a concurrent expression. But there is a
strongest and important result that relates concurrent and regular
languages.

\begin{theorem}[\cite{Shaw1978}, Relating Concurrent and Regular Expressions]

If S is a concurrent expression that does not makes use of
$\otimes$ operator, then $L_{\textsf{CE}}(S)$ is regular.

\label{FEandRE}
\end{theorem}

Another important result gives bounds to the expressiveness of
concurrent expressions.

\begin{theorem}[\cite{Gischer1981}, Bounds to Concurrent Expressions Expressiveness]

If $S$ is a concurrent expression, than $L_{\textsf{CE}}(S)$ is a
context-sensitive language.

\label{LCandLSC}
\end{theorem}

In order to increase expressivity of concurrent expressions,
allowing than to express recursively enumerable languages,
\emph{synchronized concurrent expressions}\cite{Ito1982} were
proposed. They extend concurrent expressions with synchronization
mechanisms and has been used extensively in 80's to analyse
expressiveness of synchronization mechanisms of concurrency,
mainly that based on semaphores.

\begin{definition}[Synchronized Concurrent Expressions]
\label{concsyncexpressions_def}

Be $\Sigma$ an alphabet and $\Omega$ a set of symbols that denote
synchronization primitives, where $\Sigma$ and $\Omega$ are
disjoint. A concurrent expression E over $\Sigma \cup \Omega$ is
called synchronized concurrent expression and the language
$(L_S(\Omega))$ is said to be a synchronization mechanism over
$E$. A synchronized concurrent expression over $\Sigma \cup
\Omega$, adopting the synchronization mechanism \textbf{K}, where
\textbf{K}= $L_S(\Omega)$, will be denoted by
$(E,\Sigma,\Omega,{\bf K})$.

\end{definition}

The class of synchronized concurrent expression is denoted by
\textsf{SCE}. The next definition defines the language of a
synchronized concurrent expression, which gives a meaning for the
synchronization mechanism.

\begin{definition}[Synchronized Concurrent Languages]
\label{concsynclanguages_def}

Be $(S, \Sigma, \Omega, \textbf{K})$ a synchronized concurrent
expression, where $\textbf{K} = L_S(\Omega)$. The language of $S$,
$L_{\textsf{SCE}}(S)$ is defined as below:

\begin{center}

$L_{\textsf{SCE}}(S) = \{h(x) \mid x \in L_{\textsf{CE}}(x),
\overline{h}(x) \in {\mathbf K} \}$,

\end{center}

where the homomorphisms $h$ and $\overline{h}$ are defined as
follwing:

\[ h(a) = \left\{ \begin{array}{ll}
                      a       & a \in \Sigma \\
                      \lambda & a \in \Omega \\
                  \end{array}
         \right.
\]

\[ \overline{h}(a) = \left\{ \begin{array}{ll}
                                 \lambda & a \in \Sigma \\
                                 a       & a \in \Omega \\
                             \end{array}
                     \right.
\]

\end{definition}

The class of synchronized concurrent languages that uses a
synchronization mechanism $\mathbf K$ is denoted by
$\textsf{SCE}^K$.

This definition parameterizes the adopted synchronization
mechanism. The most common are that based on semaphores, whose
most significative examples are presented in the following
paragraphs

Be $\Omega_n$ a set of synchronization primitives com $n$
s\'{\i}mbolos, where:

\begin{center}
$\Omega_n = \{ \omega_i, \sigma_i \mid i = 1, \cdots, n \}$
\end{center}

The following semaphore-based synchronization mechanisms can be
defined over $\Omega_n$\cite{Ito1982}.

\begin{enumerate}

\item{Counter semaphore: $L_S(\Omega_n) = C(n)$, where:}

$C(n) = L_{\textsf{CE}}((\sigma_1 \cdot \omega_1 +
\sigma_1))^\otimes \odot (\sigma_2 \cdot \omega_2 +
sigma_2))^\otimes \odot \cdots \odot (\sigma_n \cdot \omega_n +
\sigma_n))^\otimes)$

${\mathbf C} = \{C(n) \mid n \geq 0, where C(0) = {\lambda} \}$

\item{[0]-counter semaphore: $L_S(\Omega_n) = C_0(n)$, where:}

$C_0(n) = L_{\textsf{CE}}((\sigma_1 \cdot \omega_1))^\otimes \odot
(\sigma_2 \cdot \omega_2))^\otimes \odot \cdots \odot (\sigma_n
\cdot \omega_n))^\otimes)$

${\mathbf C_0} = \{C_0(n) \mid n \geq 0, where C_0(0) = {\lambda}
\}$

\item{Binary semaphore: $L_S(\Omega_n) = B(n)$, where:}

$B(n) = L_{\textsf{CE}}((\sigma_1 \cdot \omega_1 + \sigma_1))^*
\odot (\sigma_2 \cdot \omega_2 + sigma_2))^* \odot \cdots \odot
(\sigma_n \cdot \omega_n + \sigma_n))^*)$

${\mathbf B} = \{B(n) \mid n \geq 0, where B(0) = {\lambda} \}$

\item{[0]-binary semaphore: $L_S(\Omega_n) = B_0(n)$, where:}

$B_0(n) = L_{\textsf{CE}}((\sigma_1 \cdot \omega_1))^* \odot
(\sigma_2 \cdot \omega_2))^* \odot \cdots \odot (\sigma_n \cdot
\omega_n))^*)$

${\mathbf B_0} = \{B_0(n) \mid n \geq 0, where B_0(0) = {\lambda}
\}$

\end{enumerate}

$\mathbf C$, $\mathbf C_0$, $\mathbf B$, $\mathbf B_0$ denote the
sets of synchronization primitives with any number of primitives.

This notation can be generalized to any synchronization mechanism
other than semaphores. Let \textbf{CE}(\textbf{n}) be the family
of synchronization mechanisms with \emph{n} primitives. Then
$\mathbf{CE}=\bigcup{\mathbf{CE}(n)},\ n = 0 \dots n$. Thus,
$\mathbf {C}(n)$, $\mathbf{C_0}(n)$, $\mathbf{B}(n)$,
$\mathbf{B_0}(n)$ are special cases of $\mathbf{CE}(2n)$. A
concurrent expression controlled by a semaphore system will be
called as a \emph{semaphore controlled concurrent expression}.

Many important results about the expressive power of synchronized
concurrent expressions were presented\cite{}. But, in this work,
one deserve special attention. It establishes the equivalence of
\emph{synchronized regular expressions} that uses [0]-counter
semaphores as synchronization protocol to Petri nets. Synchronized
regular expressions are defined below.

\begin{definition}[Regular Expressions Controlled by Balanced Semaphores]
\label{def:regsyncexpressions_def}

Regular expressions controlled by balanced semaphores (RECBS) are
defined as [0]-counter synchronized concurrent expressions that
does not make use of $\otimes$ operator.

\end{definition}

The class of RECBS's is denoted by \textsf{RECBS}. Remember that
theorem \ref{FEandRE} guarantees that concurrent expressions
without $\otimes$ operator are equivalent to regular expressions.
However the presence of $\odot$ operator and [0]-counter
semaphores guarantees that this kind of expression may generate a
richer class of formal languages than simple regular expressions,
a fact enunciated by the following theorem.

\begin{theorem}[Equivalence of RECBS to Petri nets \cite{} ]
\label{theo:equivalence_pn_recbs}

The class of languages generated by a RECBS is
$\ell{_\lambda}{^1}$.

\label{SREandPN}
\end{theorem}

The result in theorem \ref{SREandPN} are convenient for our
purpose to make descriptive power of the \textsf{Hash} language equivalent to
descriptive power of Petri nets. The use of RECBS avoids the use
of $\otimes$ operator. Shaw \cite{Shaw1978}, when introducing flow
expressions to make software descriptions gave two interpretations
for $\otimes$: a \emph{parallel loop} and as a sequential loop
that creates a process (\emph{fork}) in each iteration. Neither
interpretation is practical in the \textsf{Hash} language, because it assumes
static parallelism.

\end{document}